\documentclass[twocolumn,prb,aps,superscriptaddress,longbibliography]{revtex4-1}

\usepackage{graphicx}
\usepackage{amsmath}
\usepackage{amssymb}
\usepackage{marvosym}
\usepackage{txfonts}
\usepackage{bm}
\usepackage{xcolor}
\usepackage{color}
\usepackage{ulem}
\usepackage{makecell}
\usepackage{multirow}
\usepackage[colorlinks=true,urlcolor=blue,anchorcolor=blue,linkcolor=blue,citecolor=blue,breaklinks=true]{hyperref}

\newcommand{\lsco}{La$_{2-x}$Sr$_x$CuO$_4$}
\newcommand{\lscozn}{La$_{2-x}$Sr$_x$Cu$_{1-y}$Zn$_y$O$_4$}
\newcommand{\tbcod}{Tl$_2$Ba$_2$CuO$_{6+\delta}$}
\newcommand{\tbco}[1]{Tl$_2$Ba$_2$CuO$_{#1}$}

\newcommand{\bscco}{Bi$_2$Sr$_2$CaCu$_2$O$_{8 + \delta}$}
\newcommand{\tc}{$T_c$}
\newcommand{\dwave}{$d$-wave}
\newcommand{\veck}{\textbf{k}}
\newcommand{\veckP}{\textbf{k}^\prime}
\newcommand{\Vkkp}{\left|V^i_{\veck,\veckP}\right|^2}

\newcommand{\FSavg}{_{\mathrm{FS}_{\veck}}}
\newcommand{\FSavgP}{_{\mathrm{FS}_{\veckP}}}

\def\k{{\bf k}}
\def\={&=&}

\definecolor{LSCORed}{RGB}{214,39,40}
\def\LSCOColor{\color{LSCORed}}

\definecolor{TlBlue}{RGB}{31, 119, 180}
\def\TlColor{\color{TlBlue}}

\begin{document}

\title{Effect of realistic out-of-plane dopant potentials on the superfluid density of overdoped cuprates}

\author{H. U. \"Ozdemir}
\affiliation{Department of Physics, Simon Fraser University, Burnaby, BC, V5A~1S6, Canada}
\author{Vivek Mishra}
\affiliation{Kavli Institute for Theoretical Sciences, University of Chinese Academy of Sciences, Beijing 100190, China}
\author{N.~R.~Lee-Hone}
\affiliation{Department of Physics, Simon Fraser University, Burnaby, BC, V5A~1S6, Canada}
\author{Xiangru Kong}
\affiliation{Center For Nanophase Materials Sciences, Oak Ridge National Laboratory, Oak Ridge, TN 37831, USA}
\author{T. Berlijn}
\affiliation{Center For Nanophase Materials Sciences, Oak Ridge National Laboratory, Oak Ridge, TN 37831, USA}
\author{D.~M.~Broun}
\affiliation{Department of Physics, Simon Fraser University, Burnaby, BC, V5A~1S6, Canada}

\author{P.~J. Hirschfeld}
\affiliation{Department of Physics, University of Florida, Gainesville FL 32611}

\begin{abstract}
Recent experimental papers on hole-doped overdoped cuprates have argued that   a series of observations showing unexpected behavior in the superconducting state imply the breakdown of the quasiparticle-based Landau--BCS paradigm in that doping range.  In contrast, some of the present authors have argued that a phenomenological ``dirty $d$-wave" theoretical analysis explains essentially all aspects of thermodynamic and transport properties in the superconducting state, provided the unusual effects of weak, out-of-plane dopant impurities are properly accounted for.  Here we attempt to place this theory on a more quantitative basis by performing  \textit{ab-initio} calculations of dopant impurity potentials for LSCO and Tl-2201.  These potentials are more complex than the pointlike impurity models considered previously, and require calculation of forward scattering corrections to transport properties. Including realistic, ARPES-derived bandstructures, Fermi liquid renormalizations, and vertex corrections, we show that the theory can explain semiquantitatively the unusual superfluid density measurements of the two most studied overdoped materials.
\end{abstract}

\maketitle{}

\section{Introduction}
Cuprates have represented a challenge to the central paradigms of condensed matter physics since their discovery in 1986.\cite{Keimer2015,Proust2019}  High-temperature superconductivity evolves  with doping out of the Mott state together with a host of exotic and poorly understood phenomena such as the pseudogap phase, the strange metal phase and various intertwined orders, so it has been natural to formulate the problem as one of understanding  the ground state of the  doped Mott insulator.  Indeed, there is general agreement that aspects of the underdoped phase diagram imply a definitive breakdown of the Landau--BCS quasiparticle-based approach to  interacting electrons.    On the other hand, an alternative philosophy consists in assuming that it is equally valid to attack the superconductivity problem from the overdoped side, where intertwined orders are largely absent and there is no pseudogap.   
It is assumed in such an approach that the Landau--BCS theory applies for sufficiently high doping $p$, implying that in the range of experimental interest between $p\simeq 20-30$\% it should be a good starting point, with significant but perturbative corrections due to reduced quasiparticle weights, and residual quasiparticle interactions represented by Landau parameters.

\begin{figure}[t]
    \centering
    \includegraphics[angle=0,width=0.65\linewidth]{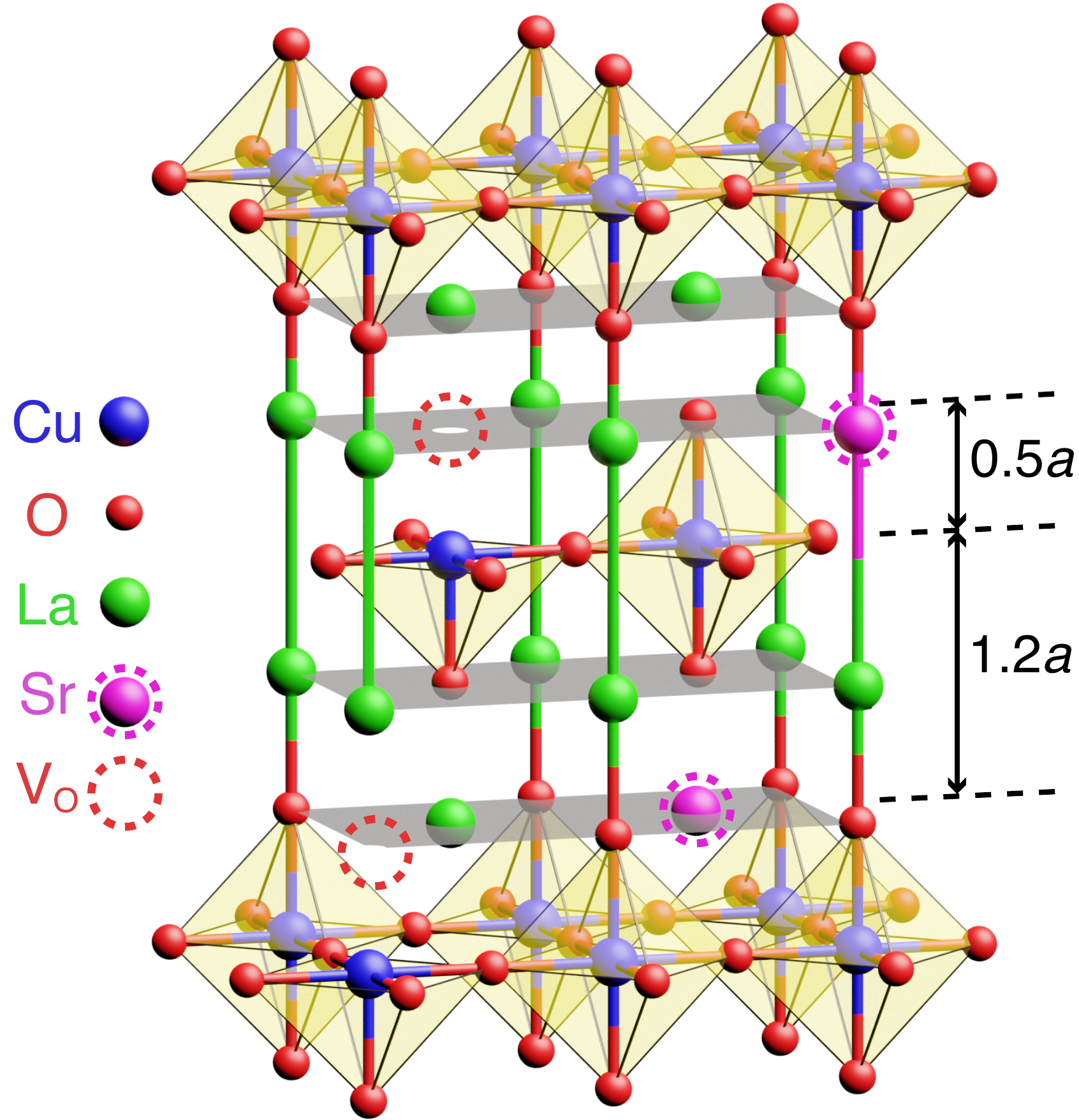}
    \caption{Several unit cells of body-centered-tetragonal La$_{2-x}$Sr$_x$CuO$_4$, showing the location of the dominant defects: Sr dopants, substituting for La at the A site; and apical-oxygen vacancies, V$_\mathrm{O}$.  (Impurity locations are indicated by dashed circles.) Note that, as experienced by electrons propagating in the CuO$_2$ planes, each defect plays a dual role: once for the closest CuO$_2$ plane, located a distance $0.5 a$ away; and again for the more distant CuO$_2$ plane, a distance $1.2 a$ away.  The body-centered tetragonal structure means that in one case the impurity is site centered, and in the other plaquette centered.  In particular, the apical-oxygen vacancy is centered directly above a Cu site, imparting significant pointlike character to its impurity potential, as we will see below.}
    \label{fig:impuritiesLSCO}
\end{figure}

\begin{figure}[t]
    \centering
    \includegraphics[angle=0,width=0.65\linewidth]{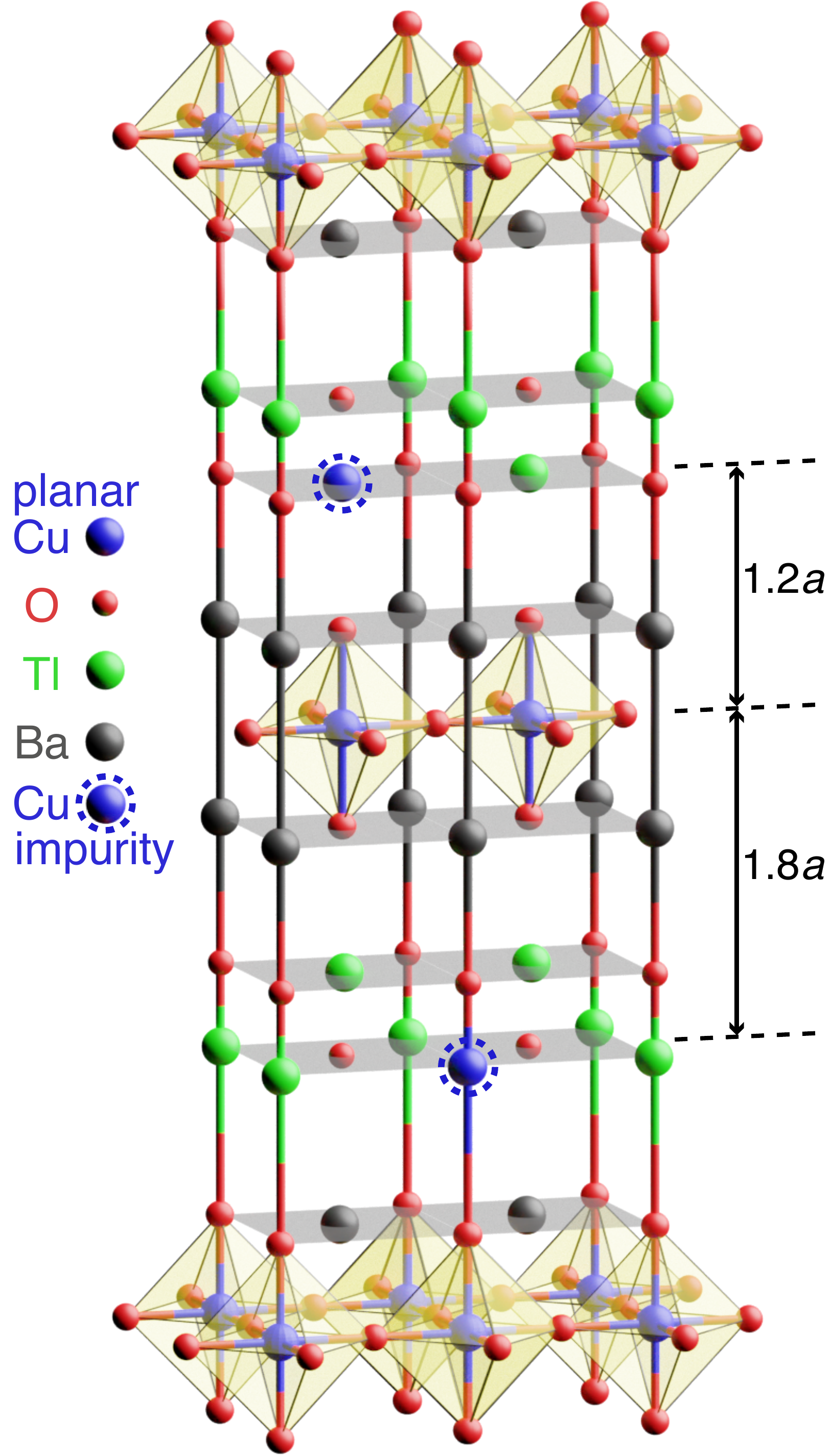}
    \caption{Several unit cells of body-centered-tetragonal Tl$_2$Ba$_2$CuO$_{4+\delta}$. Tetragonal Tl-2201 naturally grows with an excess of Cu (indicated by dashed circles) that substitutes for Tl atoms in the Tl$_2$O$_2$ double layers, and makes a significant contribution to the hole doping of this material.\cite{Peets:2010p2131} As in LSCO, these copper defects play a dual role: simultaneously site-centered with respect to the nearest CuO$_2$ plane and plaquette-centered with respect to the more distant one. (The Tl$_2$O$_2$ double layers also host interstitial oxygen dopants, not shown here.)  The large unit cell height of Tl-2201 means that these defects are located well away from the CuO$_2$ planes (at least $1.2 a$), leading to weaker impurity potentials that have strong forward-scattering character.}
    \label{fig:impuritiesTl2201}
\end{figure}

Recently, this notion was challenged by several experiments on epitaxially grown overdoped LSCO films, closely spaced in doping.  Papers reporting both superfluid density\cite{Bozovic:2016ei} and terahertz conductivity\cite{Mahmood:2017} measurements  revealed strong deviations from naive expectations for a clean $d$-wave superconductor, and claimed furthermore that disorder could not explain these effects, primarily based on the observed linearity of the measured superfluid density $\Delta\rho_s(T)$.  However, in a series of papers,\cite{Lee-Hone:2017,Lee-Hone:2018,LeeHone2020} some of the present authors argued that accounting for the weak-scattering nature of the dominant out-of-plane dopant impurities, as well as the realistic low-energy electronic structure, could explain these observations.  These works calculated, for the same set of phenomenological impurity parameters within the so-called ``dirty $d$-wave" theory,\cite{Hirschfeld:1993cka,Arberg:1993cu} not only superfluid density and optical conductivity, but also specific heat and thermal conductivity, concluding that the theory accounted well for the properties of the superconducting state in the overdoped regime of LSCO and Tl-2201 (crystal structures shown in Figs.~\ref{fig:impuritiesLSCO} and \ref{fig:impuritiesTl2201}), the two materials that have been systematically studied at high overdoping.  Nevertheless, there has been an ongoing reluctance to accept that simple impurity models, based on pointlike defects and predominantly Born-limit scatterers, are applicable and operating in a physically relevant parameter regime.\cite{Bozovic:2018dx,Mahmood:2017}

\begin{figure}[t]
    \centering
    \includegraphics[angle=0,width=\linewidth]{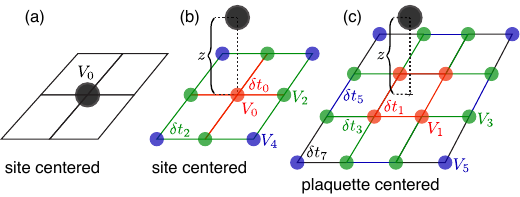}
    \caption{A one-band model of the square CuO$_2$ lattice, depicting the  tight-binding impurity potential as experienced by an electron propagating in the CuO$_2$ plane.  Terms $\delta H^i_{\bf RR'}$in the impurity Hamiltonian, Eq.~(\ref{eqn:Himp}), that are equivalent by symmetry have been grouped, resulting in a sequence of site energies, $V_i$, and a hierarchy of modifications of the nearest-neighbor hopping integral, $\delta t_i$, which are listed in \mbox{Tables~\ref{tab:impuritiesAOV}--\ref{tab:impuritiesTl} for the various impurities.} (For clarity, modifications of the next-nearest-neighbor hopping integrals are not shown, but are also included in the model.)  (a)~An in-plane, site-centered defect, e.g., a Cu vacancy in the CuO$_2$ plane, which we treat as a point scatterer with a single, on-site impurity term $V_0$; (b) an out-of-plane, site-centered defect, e.g., an apical O vacancy, located a height $z$ above the nearest Cu; and (c) an    out-of-plane, plaquette-centered defect, e.g., a Sr substituting for La at an A site located a height $z$ above the nearest CuO$_2$ plaquette. Note that while a site-centered defect can be predominantly pointlike (i.e., $V_0 \gg V_2,V_4,...$), a plaquette-centered defect is always equidistant from its four closest Cu sites: it therefore has finite range and is inherently of forward-scattering character.}
    \label{fig:potential_schematic}
\end{figure}

Accounting semiquantitatively for differences among overdoped cuprate materials requires a detailed understanding of the dopant impurities themselves.   In Refs.~\onlinecite{Lee-Hone:2017,Lee-Hone:2018,LeeHone2020},  overdoped cuprates were modeled by a 2D tight-binding band, and specifying a relatively weak onsite impurity potential, of order 0.1$t$, where $t$ is the nearest neighbor Cu--Cu hopping in the plane.  But
in real cuprates, the dopants and defect atoms are situated in
various sites of the crystal, and may therefore be expected to produce quite different effective scattering potentials as experienced by electrons propagating in the CuO$_2$ plane.  For example, in LSCO, both the apical oxygen vacancies, and the Sr dopants that substitute for La at the so-called A site, are located only one layer away ($0.5 a$ away, where $a$ is the in-plane lattice spacing) from the nearest CuO$_2$ plane (Fig.~\ref{fig:impuritiesLSCO}). In \mbox{Tl-2201}, oxygen interstitials, and the excess Cu that substitutes for Tl,\cite{Peets:2010p2131} are located at least two layers ($1.2 a$) away, producing 
correspondingly weaker in-plane potentials\cite{Fujita2005,Rullier-Albenque2008} (Fig.~\ref{fig:impuritiesTl2201}).  In addition to the effect on the magnitude of the potential, defects residing outside the CuO$_2$ planes produce a longer-range scattering ``footprint" in the CuO$_2$ plane, which is particularly important for transport, where a predominance of forward scattering enhances the current.  As we show below, the in-plane momentum dependence of the effective impurity potential is nontrivial, being determined by the projection of the impurity position onto the CuO$_2$ plane.  In particular, the nature of the scattering depends sensitively on whether the defect is site-centered or plaquette-centered with respect to the CuO$_2$ lattice, as shown schematically in Fig.~\ref{fig:potential_schematic}. (For the body-centered tetragonal structures shown in Figs.~\ref{fig:impuritiesLSCO} and \ref{fig:impuritiesTl2201}, each defect is simultaneously site-centered with respect to one neighboring CuO$_2$ plane, and plaquette-centred --- albeit at different distance --- from its other neighboring CuO$_2$ plane.)  These factors all influence the strength and range of the potential, and in turn determine the degree of pairbreaking and the resulting anisotropy of the quasiparticle and transport scattering rates.

Recognizing the importance of apical oxygen vacancies in LSCO, a recent study has taken the first steps toward understanding the role of these defects in the overdoped cuprates.\cite{Wang2022} Unfortunately, that work assumed an incorrect form for the impurity potential, stating without proof that the apical O vacancy has zero potential on the Cu site below it. In addition, the study did not carry out a self-consistent treatment of the self energy, using instead the normal-state form for the scattering rate.  Furthermore, Ref.~\onlinecite{Wang2022} neglected to consider forward-scattering corrections and explicit gap renormalization, both of which are important when the impurity potential is momentum dependent, and
 have a significant effect on both the qualitative and quantitative behavior of the superconductivity and the electrodynamics. We discuss this work further in a separate comment.\cite{comment}

Confirming the general dirty $d$-wave description of the overdoped cuprates thus requires a more accurate calculation of impurity potentials, including  the vertex corrections to the transport properties that arise with  extended impurity potentials.  In this paper, we perform a series of \textit{ab-initio} calculations of the tight-binding impurity potentials due to various dopants and defects, including their local modification of the hopping parameters, using a Wannier-function-based supercell approach,\cite{Berlijn2011} and revisit the calculations of Refs.~\onlinecite{Lee-Hone:2017,Lee-Hone:2018,LeeHone2020}.  We also discuss the influence of  band structure effects, including the proximity in the overdoped regime of the van Hove singularity at the $(\pi,0)$ point, which necessitates the use of momentum sums rather than Fermi-surface integrals. We find that  the success of the previous phenomenological approach  is confirmed by the more complete \textit{ab-initio}-based calculations presented here.  
    
The nonuniversal details of the calculations may seem tedious to those interested in drawing broad conclusions for the generic cuprate phase diagram, but we regard them as essential to establishing the applicability of the quasiparticle-based Landau--BCS paradigm to the overdoped cuprates in the sense described above.  We believe that it is important to  investigate conventional explanations  for the apparently unusual physics uncovered in Refs.~\onlinecite{Bozovic:2016ei,Mahmood:2017}  and other works before turning to more exotic explanations.\cite{Mazin2022}  Conversely, if alternative theories beyond the Landau--BCS paradigm are put forward, it seems reasonable to insist that they be capable of explaining the experimental data, and the variation of properties between different cuprates, at the same level of detail as presented here.

    \section{Formalism}
    \subsection{Band structure}
\label{subsec:dopant_potential}
    For our purposes here, an \textit{ab-initio} description of the effective impurity potentials is desired to convince the reader that a realistic, materials-specific description of overdoped cuprates has been obtained.  Here one immediately encounters the usual challenge associated with describing cuprates by first principles approaches, namely that density functional theory (DFT) in its simplest form fails for the cuprate parent compounds because it cannot describe the Mott insulating state.
    While considerable progress has been made incorporating correlations into DFT to describe high-energy physics,\cite{Kent2018,Furness2018} these methods are not appropriate for a complete \textit{ab-initio} description of low-energy phenomena, accurate at the few meV level.  We therefore adopt a hybrid approach, starting with a low-energy tight-binding model describing the CuO$_2$-plane states of the notional pure material, with hopping matrix elements taken from fits to ARPES measurements on LSCO\cite{Yoshida:2006hw,Yoshida:2007} and Tl-2201.\cite{Plate:2005}   To capture the continuous evolution of electronic structure with doping,  tight-binding parameters for LSCO are interpolated between the fixed dopings at which the ARPES measurements were performed.\cite{Yoshida:2006hw}  The only parameter with strong doping dependence is the chemical potential, which is set using the direct correspondence between hole doping and Fermi volume. The doping dependence of the next-nearest-neighbour hopping is relatively weak, and all other tight-binding parameters are doping independent. For Tl-2201, limited ARPES data\cite{Plate:2005} means that doping dependence of electronic structure is generated via rigid band-shift, with the only doping-dependent parameter the chemical potential. (For more details of the procedure see Refs.~\onlinecite{Lee-Hone:2017,LeeHone2020}.)
 
 \begin{figure}[t]
    \centering
        \includegraphics[width=0.7 \columnwidth,scale=1.0]{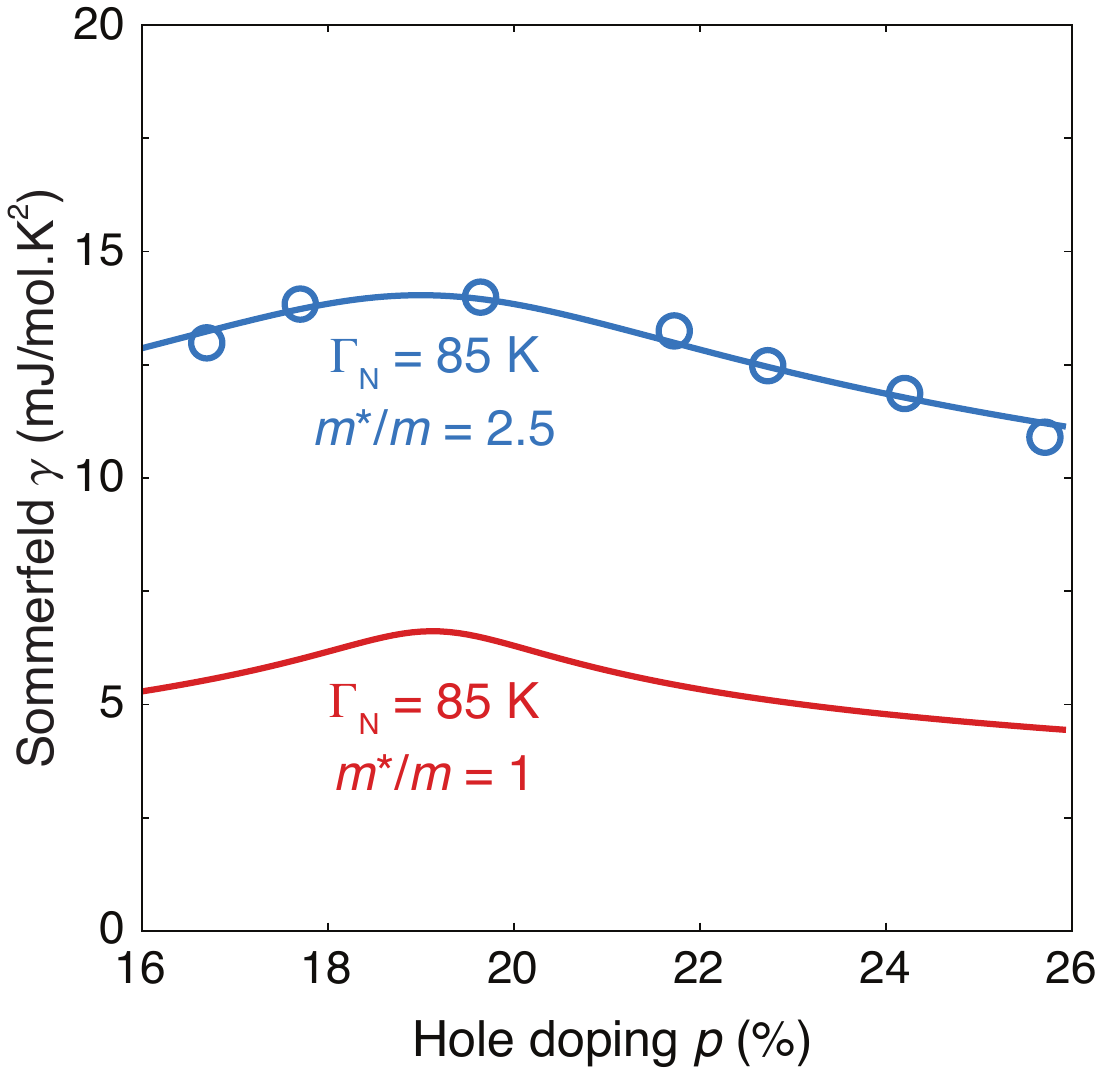}
    \caption{Sommerfeld specific heat coefficient $\gamma$ for Zn-doped LSCO.  Experimental heat capacity data from Ref.~\onlinecite{Momono:1994et} are fit using the ARPES-derived bandstructure\cite{Yoshida:2006hw} with scattering rate $\Gamma_N$ and Fermi-liquid mass enhancement $m^\ast\!/m$ as adjustable parameters, from which we determine $m^\ast\!/m = 2.5$. The unrenormalized Sommerfeld coefficient is shown below for comparison.}
    \label{fig:Sommerfeld}
\end{figure}   
 
    The only additional renormalization that needs to be considered when modeling the pure electronic structure is the many-body renormalization that occurs close to the Fermi level, which flattens the dispersion $\xi_\veck$ by a factor of $m/m^\ast$.  As discussed in Ref.~\onlinecite{LeeHone2020}, the ARPES measurements of Plate~\textit{et~al}.\ for Tl-2201 were performed at very low energies\cite{Plate:2005} \mbox{(tens of meV)} and already capture the $m^\ast$ renormalization.  In LSCO, by contrast, the ARPES tight-binding fits to the dispersion\cite{Yoshida:2007} are carried out over a wider energy range --- typically 0.5~eV --- and therefore do not include the many-body renormalization.  While some flattening of the LSCO dispersion is visible in ARPES (see Fig.~6 of Ref.~\onlinecite{Yoshida:2007}), it is not sufficiently well resolved to determine $m^\ast/m$. Instead, we turn to heat capacity measurements on Zn-doped LSCO,\cite{Momono:1994et,Wang2007} where the Zn doping has been used to suppress $T_c$ in order to access the Sommerfeld coefficient in the low temperature limit.  We model the Zn dopants as strong-scattering impurities with normal-state scattering rate $\Gamma_N$, and obtain a good fit to the doping-dependent Sommerfeld coefficient with $\Gamma_N = 85$~K and $m^\ast/m = 2.5$, as shown in Fig.~\ref{fig:Sommerfeld}.  This value of $m^\ast/m$ has been used in all subsequent calculations on LSCO.

\begin{figure*}[t]
    \centering
        \includegraphics[width=
        1.0\linewidth,scale=1.0]{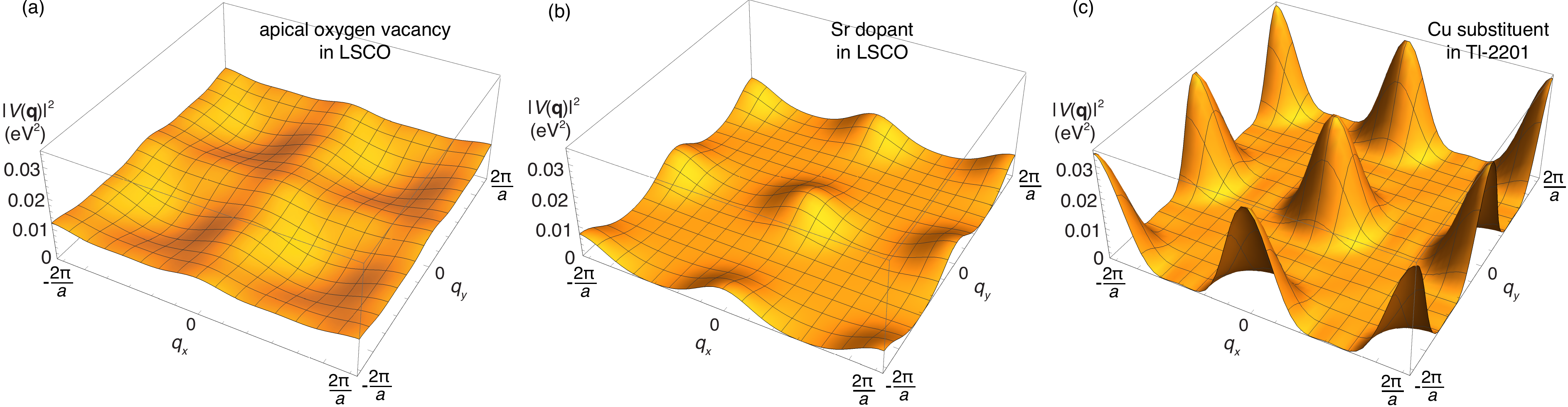}
    \caption{Impurity matrix elements as a function of momentum transfer $\veck - \veckP \equiv \mathbf{q} = (q_x,q_y)$ for: (a) the apical oxygen vacancy in LSCO; (b) the Sr dopant in LSCO; and (c) the Cu--Tl substitution in Tl-2201. Matrix elements $V(\mathbf{q})$ are obtained from Eq.~\ref{eqn:matrixelements}, using the site-energies $V_\mathbf{R}^i$ from Tables~\ref{tab:impuritiesAOV} to \ref{tab:impuritiesTl}, respectively. (Hopping modifications are not included, but are in any case small.)}
    \label{fig:impurity_potentials}
\end{figure*}      

   \subsection{Impurity potentials}
    We then describe the influence of the impurity in question in terms of a lattice impurity Hamiltonian $H_\mathrm{imp}$, as described below. The impurity Hamiltonian consists of modifications of both the site energies and hopping matrix elements relative to the pure system, and is determined by a Wannier-based method for calculating impurity potentials\cite{Berlijn2011} that has been used successfully, e.g., to accurately simulate strong disorder in superconducting materials.\cite{Berlijn2011,Berlijn2012,Wang:2013bq,Berlijn:2014fb}
    In this method, two DFT calculations are carried out for each type of impurity: one of a $3 \times 3 \times 1$ supercell containing a single impurity (La$_{35}$SrCu$_{18}$O$_{72}$, La$_{36}$Cu$_{18}$O$_{71}$ or Tl$_{35}$Ba$_{36}$Cu$_{19}$O$_{108}$); and a reference calculation for the corresponding pure system (La$_4$Cu$_2$O$_8$ or Tl$_4$Ba$_4$Cu$_2$O$_{12}$).  For the (La,Sr) substitution and the O vacancy impurity potential in LSCO we have repeated the calculations for two different hole dopings, determined in each case by integrating the density of states (DOS) down from the top of the band to the Fermi level. In addition to the undoped calculations (in which no holes or electrons were added, other than those added by the impurity itself), we have repeated the calculations with a hole doping ($p_\mathrm{vH}$) for which the van Hove singularity in the single impurity supercell lies at the Fermi level, and a second doping with 0.07 extra holes per Cu ($p_\mathrm{vH}+7\%$). The reason for deriving the potentials at $p_\mathrm{vH}$ and $p_\mathrm{vH} + 7\%$ is that the enhanced density of states at the van Hove singularity has a strong effect on screening, making the impurity potentials substantially more local. More details of the first principles derivation of the impurity potential are given in Appendix~\ref{appendix_impurities}.

    Each pair of DFT calculations is Wannier-projected onto a one-orbital lattice to obtain the supercell Hamiltonian for the $i^\mathrm{th}$ impurity type, $H_\mathrm{supercell}^i$, and the reference Hamiltonian, $H_0$, respectively.   The difference between the two tight-binding models then defines the Hamiltonian associated with the $i^\mathrm{th}$ impurity,
    \begin{equation}
    \label{eqn:Himp}
    \begin{split}
        H_\mathrm{imp}^i & \equiv \left(H_\mathrm{supercell}^i - \mu^i  {\hat N^i}\right) - \left(H_0 - \mu_0 {\hat N}\right)\\
        & = \sum_{\mathbf{R},\mathbf{R}^\prime\!,\sigma}\!\!\delta H_{\mathbf{R}\mathbf{R}^\prime}^i c_{\mathbf{R}\sigma}^\dagger c_{\mathbf{R}^\prime\sigma}\\
        & \equiv \sum_{\mathbf{R}, \sigma} V_\mathbf{R}^i c_{\mathbf{R},\sigma}^\dagger c_{\mathbf{R}\sigma} + \!\! \sum_{\mathbf{R}\ne\mathbf{R}^\prime\!,\sigma}\!\!\delta t_{\mathbf{R}\mathbf{R}^\prime}^i c_{\mathbf{R}\sigma}^\dagger c_{\mathbf{R}^\prime\sigma}\;,
    \end{split}
    \end{equation}
    for which the Bloch-wave matrix elements are
    \begin{equation}
    \label{eqn:matrixelements}
    \begin{split}
    V^i_{\veck,\veckP} &= \sum_{\mathbf{R},\mathbf{R}^\prime} \delta H_{\mathbf{R}\mathbf{R}^\prime}^i e^{-i \veck \cdot \mathbf{R}} e^{i \veckP \cdot \mathbf{R}^\prime}\\
    &= \sum_\mathbf{R} V_\mathbf{R}^i e^{-i(\veck - \veckP)\cdot \mathbf{R}} + \sum_{\mathbf{R}\ne\mathbf{R}^\prime} \delta t_{\mathbf{R}\mathbf{R}^\prime}^i e^{-i \veck \cdot \mathbf{R}} e^{i \veckP \cdot \mathbf{R}^\prime}.
    \end{split}
\end{equation}
Here $\mu^i$ and $\mu_0$ are the chemical potentials of the simulation with and without the impurity, respectively,  determined directly from the DFT calculations via integration of the DOS.  The 2D lattice vectors $\mathbf{R}$ and $\mathbf{R}^\prime$ are measured in a coordinate system in which the impurity sits directly above (or below) the origin. Unlike the tight-binding Hamiltonian of a translationally invariant system, where all the site energies $V_\mathbf{R}$ are equivalent and the hopping integrals $t_{\mathbf{R}\mathbf{R}'}$ can be classified into a small hierarchy of increasing-range terms (nearest neighbor, next-nearest neighbor, etc.), the impurity Hamiltonian lacks translational symmetry, with $V_\mathbf{R}^i$ and $\delta t_{\mathbf{R}\mathbf{R}^\prime}^i$ falling off with distance from the impurity site.  It is nevertheless useful to group equivalent terms together: this is illustrated in Fig.~\ref{fig:potential_schematic}, and leads to the momentum-dependent form factors discussed in Appendix~\ref{appendix_impurities} and listed in \mbox{Tables~\ref{tab:impuritiesAOV}--\ref{tab:impuritiesTl}}.  For the materials of interest here, defects can be classified into two types ––– site-centered and plaquette-centered --- depending on where they sit with respect to the Cu atoms of the CuO$_2$ plane.
For the body-centered-tetragonal structures of LSCO and Tl-2201 shown in Figs.~\ref{fig:impuritiesLSCO} and \ref{fig:impuritiesTl2201}, each out-of-plane defect in fact plays a dual role: site-centered with respect to one neighboring CuO$_2$ plane and plaquette-centered with respect to the other.  To illustrate the qualitative differences between the impurity potentials, the impurity matrix elements are plotted in Fig.~\ref{fig:impurity_potentials} as $|V(\mathbf{q})|^2$, using the site energies from \mbox{Tables~\ref{tab:impuritiesAOV}--\ref{tab:impuritiesTl}} and Eq.~\ref{eqn:matrixelements}.  The Cu substituents in Tl-2201, located well away from the CuO$_2$ planes, have strong forward-scattering character, with $|V(\mathbf{q})|^2$ peaked sharply near $q = 0$ and its Umklapp replicas.  By contrast, the impurity matrix elements in LSCO are more rounded, reflecting the much smaller separation of the defects from the CuO$_2$ planes.  The pointlike nature of the apical oxygen vacancy results in a significant constant contribution to $|V(\mathbf{q})|^2$.

Since DFT calculations on cuprates systematically overestimate the electronic bandwidth relative to the bandwidth measured in ARPES experiments, the DFT impurity potentials are first expressed in units of the DFT-derived nearest-neighbor hopping $|t|$, with the physically relevant value of $|t|$ set later by ARPES, when including the impurity potentials in subsequent parts of the calculation. With these approximations in mind, a full, material-specific tight-binding model including disorder is obtained, to which we add a phenomenological \mbox{$d$-wave} pair potential to describe the superconducting state.  The same procedure is followed systematically for all material systems considered.
    
As mentioned above, the dominant defects in LSCO are apical oxygen vacancies and the Sr$^{2+}$ dopant ions that substitute for La$^{3+}$, thereby removing electrons from the conduction band.  It is expected that each added Sr atom should dope one hole into the CuO$_2$ planes, but the reality is more complicated. A discrepancy between Fermi volume and Sr content has been found in a number of ARPES studies,\cite{Yoshida:2006hw,Chang:2013ev,Horio:2018} with the careful 3D Fermi-surface measurements of Ref.~\onlinecite{Horio:2018} revealing a Fermi volume equivalent to $p = 0.32$, significantly larger than the Sr concentration of their La$_{1.78}$Sr$_{0.22}$CuO$_4$ sample, $x = 0.22$.  A possible explanation of this effect is that in chemically tuning the Sr concentration, the equilibrium oxygen content is also changed.  One way of taking this into account is to set the doping-dependent concentration of Sr defects, $n_\mathrm{Sr}$, using the relation \mbox{$n_\mathrm{Sr}=x=0.69p$}. We also consider the conventional relation, \mbox{$n_\mathrm{Sr}= p$}, and show that this does not significantly change our results.  The relevant concentration of oxygen vacancies is more difficult to determine, but is known to be significant.\cite{Torrance:1988iz,Higashi1991,Kim:2017tk} There are two inequivalent oxygen sites in LSCO: planar oxygen and apical oxygen. Structural refinements of \mbox{x-ray} diffraction data for LSCO in Ref.~\onlinecite{Higashi1991} reveal that the planar oxygen site is fully occupied in well-annealed crystals (i.e., annealed at 500$^\circ$C for 1~week, in 1~atm~O$_2$).  By contrast, the apical oxygen is much less tightly bound, with the same sample containing apical oxygen vacancies at the $n_{\mathrm{V}_\mathrm{O}} \approx 9\%$ level.  Apical oxygen vacancies are also likely to be highly relevant to the molecular-beam epitaxy (MBE) thin films\cite{Bozovic:2016ei} of interest in the current study: their large  lateral dimension makes diffusion times long, and their very high quality and crystallinity mean there are no grain boundaries or screw dislocations to provide an easy diffusion path perpendicular to the film.  This is particularly apparent when the MBE films are patterned into narrow microbridges and annealed in ozone\cite{Bozovic:2016ei} ––– rapid diffusion of oxygen across the microbridges reduces their residual resistivity by up to a factor of 4 compared to the unpatterned cm$^2$ films.\cite{Mahmood:2017} With this in mind, our calculations of superfluid density for LSCO have been performed over a range of apical oxygen vacancy concentrations, $4\% \le n_{\mathrm{V}_\mathrm{O}} \le 8\%$.

The dominant defect in Tl-2201 is a consequence of its crystal growth.  The high volatility of Tl$_2$O$_3$ at the growth temperature leads to a deficit of Tl, which is replaced by an excess of Cu that substitutes onto 4\% to 7.5\% of Tl sites.\cite{Liu:1992jx,Kolesnikov:1992gg,Hasegawa:2001bt,Peets:2010p2131}  Attempts to suppress the Cu--Tl substitution using high-pressure encapsulation\cite{Peets:2010p2131} have been partially successful, but reveal that the excess Cu$^+$ plays a vital role in making Tl-2201 an overdoped cuprate.  (Cu$^+$ has a valence of -2 relative to Tl$^{3+}$, making it an effective hole dopant. Interstitial O$^{2-}$ in the Tl$_2$O$_2$ double layers is similarly effective at doping holes into the CuO$_2$ planes.) We have calculated the impurity potential of the Cu substituent and, assuming each Cu dopes two holes, have set its  concentration to be $n_\mathrm{Cu} = p/2$. Here $n_\mathrm{Cu}$ gives the concentration of Cu substituents as a fraction of the in-plane Cu and varies smoothly from 8\% to 15\% across the overdoped range. This is broadly in line with the direct measurements of Cu concentration cited above, remembering that there are two Tl sites for every in-plane Cu. To the extent that oxygen interstitials are present, we expect them to  behave as dopants, and as scatterers, in much the same way as the Cu substituents, given their similar location and relative valence.
    
\subsection{Superconducting State}
Within the Matsubara formalism, the Nambu space Green's function for a dirty superconductor can be written as\cite{Hirschfeld:1994}
\begin{equation}
{\underline G}(\veck,i \omega_n)=- \frac{i \tilde\omega_{\veck,n} \tau_0 + \tilde\Delta_{\veck,n} \tau_1 + \xi_\veck \tau_3}{\tilde\omega_{\veck,n}^2 + \tilde\Delta_{\veck,n}^2 + \xi_\veck^2}\;,
\end{equation}
where $\xi_\veck$ is the single-particle dispersion relative to the Fermi level and $\tau_i$ are the Pauli
particle--hole matrices. In anticipation of momentum-dependent scattering effects that arise from out-of-plane scatterers (and unlike the more common case of point scatterers) we include momentum dependence in the renormalized Matsubara frequencies, $\tilde\omega_{\veck,n}\equiv\omega -\Sigma_0(\k,\omega_n)$, and allow for explicit renormalization of the superconducting gap, $\tilde \Delta_{\veck,n}\equiv \Delta_\k+\Sigma_1(\k,\omega_n)$.  We give the explicit forms of $\tilde\omega_{\veck,n}$ and $\tilde\Delta_{\veck,n}$ below.  In principle, the band energy $\xi_\k$ should also be self-consistently renormalized by the disorder, but this effect is generically weaker, indeed vanishing identically in the case of pointlike impurities in a particle--hole symmetric system.  Furthermore, the real part of the self energy $\Sigma_3\equiv \tilde\xi_\k-\xi_\k$ is already incorporated into the ARPES-derived electronic structure employed here. 

\begin{figure}[t]
    \centering
        \includegraphics[width=\linewidth,scale=1.0]{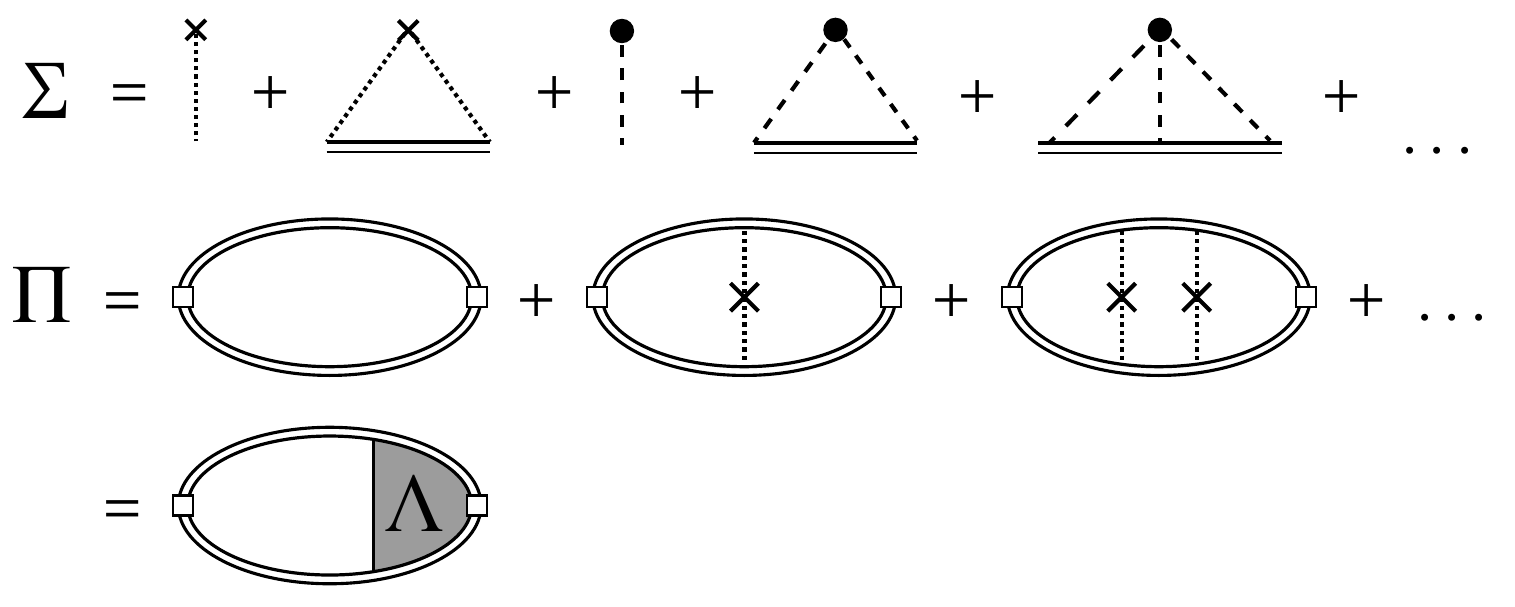}
    \caption{Diagrammatic expansions of the disorder-averaged self energy $\Sigma$ and polarization bubble $\Pi$, in the realistic disorder model. Circles denote  pointlike strong impurities, which are treated in the $t$-matrix approximation. Crosses denote one type of weak extended scatterer, treated in the Born approximation. For simplicity the other species of weak scatterer are not shown, but appear as additive terms in the actual calculations. Only weak scatterers contribute to the vertex function $\Lambda$, and there are no crossing diagrams between different impurity types.}
    \label{fig:diagrams}
\end{figure} 

For a separable pairing interaction, $V_0 d_\veck d_{\veckP}$, the gap equation for a weak-coupling unconventional superconductor is\cite{Hirschfeld:1994}  
\begin{equation}
\Delta_{\veck} = \frac{2 T}{N} \sum_{\omega_{n>0}}^{\Omega_c}\sum_{\veckP} V_0 d_{\veck} d_{\veckP} \frac{\tilde \Delta_{\veckP\!\!,n}}{\tilde \omega_{\veckP\!\!,n}^2+\tilde \Delta_{\veckP\!\!,n}^2+\xi_{\veckP}^2 }\;,
\end{equation}
where $\Omega_c$ is a high frequency cutoff, the $\veckP$ sum runs over the first Brillouin zone, and $N$ is the number of sites in the lattice. For the tetragonal cuprates we are interested in, the $d$-wave eigenfunction takes the form
\begin{equation}
    d_\veck \propto \left[\cos(k_x a) - \cos(k_y a)\right]\;,
\end{equation}
where $a$ is the in-plane lattice spacing, and $d_\veck$ is normalized such that \mbox{$\frac{1}{N}\sum_{\veck}d_\veck^2 = 1$}.  The system thus condenses into the state $\Delta_\k=\Delta_0 d_\veck$ at all temperatures below $T_c$.

In addition to the weak-scattering \textit{ab-initio} potentials  discussed above for the out-of-plane impurities, we also allow for a small density of in-plane strong scatterers, such as Cu vacancies.  The combined effect of the strong and weak scatterers is shown diagrammatically in Fig.~\ref{fig:diagrams}.  Since the out-of-plane potentials are relatively weak, they can be treated in the Born approximation, a statement that we justify in more detail below. The in-plane strong scatterers are treated as pointlike unitarity scatterers in the $t$-matrix approximation.  The out-of-plane defects generate potentials with significant spatial extent, leading in turn to strong momentum dependence in the matrix elements $V_{\veck,\veckP}$.  As a result, vertex corrections must be included in the calculation of two-particle properties such as superfluid density,  resistivity and optical conductivity.

Including the extended nature of the dopant defects, the renormalization equations acquire a somewhat different form than is usual, but the self-energy contributions from the various impurity types are still additive.\cite{Nunner:2005p654}
\begin{align}
    &\tilde{\omega}_{\veck,n}=\omega_{n} + \frac{1}{N}\sum_{i,\;\veckP} n_i \Vkkp \!\!\frac{\tilde \omega_{\veckP\!\!,n}}{\tilde \omega_{\veckP\!\!,n}^2+\tilde \Delta_{\veckP\!\!,n}^2+\xi_{\veckP}^2 } + \frac{\Gamma_N^U}{G_0}   \label{eqn:tau0_self_energy}\\
    &\tilde{\Delta}_{\veck,n} = \Delta_{\veck} + \frac{1}{N}\sum_{i,\;\veckP} n_i \Vkkp \!\! \frac{\tilde \Delta_{\veckP\!\!,n}}{\tilde \omega_{\veckP\!\!,n}^2+\tilde \Delta_{\veckP\!\!,n}^2+\xi_{\veckP}^2}\;,
    \label{eqn:tau1_self_energy}
\end{align}
where 
\begin{equation}
    G_0 = \frac{1}{\pi N N_0} \sum_\veck \tfrac{1}{2} \mathrm{Tr}\left[\tau_0 {\underline G}(\veck,i \omega_n)\right]\;;
\end{equation}
$N_0$ is the single-spin density of states { per unit cell}, calculated self-consistently at the Fermi level; $n_i = N_i/N$ is the concentration of the $i^\mathrm{th}$ impurity; and $\Gamma_N^U$ is the normal-state scattering rate associated with the strong-scattering impurities. 

 We note that the use of the Born approximation has been criticized as corresponding to an unphysical limit consisting of an infinite number of infinitely weak scatterers.\cite{Mahmood:2017} This is too narrow an interpretation. In fact, for dilute pointlike scatterers in effectively three dimensions, the Born approximation is justified whenever the quantity $V_\mathrm{imp}N_0$ is significantly less than one.  (It is also common to parameterize the strength of a scatterer in terms of the cotangent of its scattering phase shift, $c = \cot \delta = 1/(\pi V_\mathrm{imp} N_0)$: the Born-limit then corresponds to $c$ significantly greater than 1.) In our previous work assuming pointlike impurities, we compared results for superfluid density calculated over a wide range of scattering phase shifts and showed that $\rho_s(T)$ was qualitatively unchanged for $c \ge 2$, illustrating the broad applicability of the Born limit.\cite{LeeHone2020} For the extended impurities considered here, the need to include vertex corrections to two-particle properties would make the full $t$-matrix approximation  computationally expensive. There is, however, a simple test we can perform.  The \textit{ab-initio} calculations show that our strongest scatterer, the near apical oxygen vacancy in LSCO, is essentially pointlike (i.e., $V_0 \gg V_2, V_4, ...$), allowing it to be treated as a point scatterer in the $t$-matrix approximation,  then compared to results obtained in the Born approximation.  This test is carried out in Appendix~\ref{appendix_intermediate_scatterers}, with the conclusion that the Born approximation is well justified for all of our out-of-plane extended defects. 
 
 Another interesting aspect of extended impurities is that they give rise to momentum-dependent first-order corrections to the impurity self-energy, in contrast to the case of pointlike scatterers, for which the first-order correction is a constant that can be absorbed into the chemical potential. However, as discussed in Appendix~\ref{appendix_impurities}, the momentum dependence of the first-order self-energy arising from the hopping modifications corresponds to the same form factors as in the normal state dispersion, such that these terms simply renormalize the band structure.  Since we use tight-binding models derived from ARPES experiment, these effects are already incorporated into the dispersions adopted.

\subsection{ Transition temperatures  $T_{c0}$ and $T_c$}
For a clean system, $\omega_n$ and $\Delta_\veck$ are unrenormalized and the gap equation is\cite{Hirschfeld:1994}
\begin{equation}
    \Delta_\veck = \frac{2 T}{N} \sum^{\Omega_c}_{\omega_n > 0}\sum_{\veckP} V_0 d_\veck d_{\veckP} \frac{\Delta_{\veckP}}{\omega_{\veckP\!\!,n}^2  + \Delta_{\veckP}^{2} + \xi_{\veckP}^2}\;.
    \label{eq:cleangapequation}
\end{equation}
This allows the pairing strength $V_0$ and cutoff $\Omega_c$ to be parameterized in terms of a notional, clean-limit transition temperature $T_{c0}$, by carrying out the Matsubara sum in Eq.~(\ref{eq:cleangapequation}) at temperature $T_{c0}$, where $\Delta_\veck$ is vanishingly small and can be eliminated from the denominator:
\begin{equation}
    \frac{1}{V_0} = \frac{2 T_{c0}}{N} \sum^{\Omega_c}_{\omega_n>0}\sum_{\veck} d_\veck^2 \frac{1}{\omega_{\veck,n}^2 + \xi_{\veck}^2}\;.
    \label{pairing}
\end{equation}
 In the presence of disorder, the gap equation  at $T_c$ reduces to
\begin{equation}
        \Delta_\veck = \frac{2 T_c}{N} \sum^{\Omega_c}_{\omega_n > 0}\sum_{\veckP} V_0 d_\veck d_{\veckP} \frac{\tilde\Delta_{\veckP\!\!,n}}{\tilde\omega_{\veckP\!\!,n}^2 + \xi_{\veckP}^2}\;.
        \label{gap@Tc}
\end{equation}
 We now take $T_c(p)$ to be given roughly by experiment, in each case assuming a parabolic form for the ``superconducting dome".  Equations~(\ref{pairing}) and (\ref{gap@Tc}) can then be solved numerically in the presence of disorder to infer the value of $T_{c0}$ required to produce a given $T_c$. The variation of $T_{c0}$ with doping reflects  the intrinsic physics of the cuprate pairing interaction, in particular its weakening on the overdoped side.  We note that $T_{c0}$ should be considered an upper bound on the clean-limit transition temperature, as we have not included pair-breaking effects arising from inelastic scattering, or fluctuation effects such as the Berezinskii--Kosterlitz--Thouless vortex-unbinding transition.

 \begin{figure*}[t]
    \centering
        \includegraphics[width=1.0\linewidth,scale=1.0]{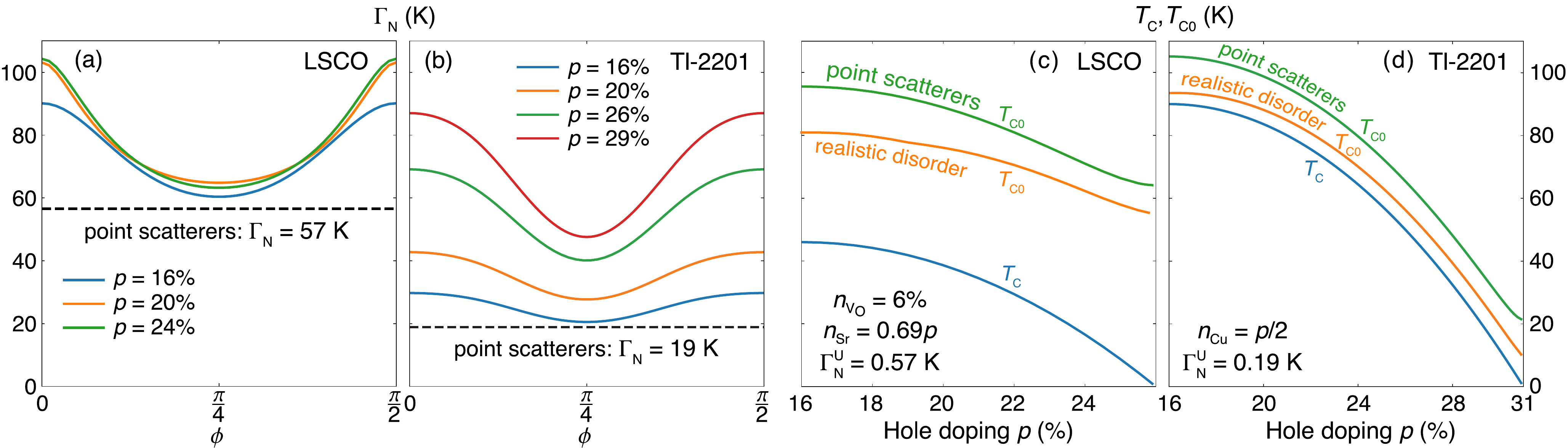}
    \caption{Comparison of point-scattering and realistic disorder models in LSCO and Tl-2201. (a,b) Normal-state elastic scattering rate $\Gamma_N$ as a function of angle $\phi$, measured from the zone axes.  Dashed lines indicate the values of $\Gamma_N$ used in the point-scattering models of Refs.~\onlinecite{Lee-Hone:2017,Lee-Hone:2018,LeeHone2020}.  (c,d) Clean-limit transition temperatures $T_{c0}(p)$ parameterize the pairing strength required to produce the parabolic ``superconducting domes" $T_c(p)$ of LSCO and Tl-2201.  Disorder parameters for the LSCO and Tl-2201 plots are as shown in panels (c) and (d), respectively.}
    \label{fig:GammaN_TcTc0}
\end{figure*}

\subsection{Superfluid density}
The superfluid density is calculated within the Matsubara formalism by evaluating the electromagnetic response function in the static limit, $\Omega \to 0$:\cite{Skalski:1964,Durst:2000}
\begin{eqnarray}
\mathcal{Q}^{ij}\= \frac{e^{2} T}{N} \sum_{\omega_n>0}\sum_\veck \textbf{v}^i_\veck \operatorname{Tr}\left[\underline{G}(\k,\omega_n) \mathbf{\Lambda}^{j} \underline{G}(\k,\omega_n)\right.\nonumber\\
&&~~~~\left.
-\tau_{3} \underline{G}(\k,\omega_n) \mathbf{\Lambda}^{j} \tau_{3} \underline{G}(\k,\omega_n)
\right]\;,
\end{eqnarray}
where the first term and second terms are the diagmagnetic and paramagnetic contributions, respectively. The group velocity along the $j$-direction is $\mathbf{v}_\veck^j = {\partial \xi_{\veck}}/{\partial k^{j}}$, and $\mathbf{\Lambda}$ is the impurity-renormalized current vertex function,
\begin{equation}
  \mathbf{\Lambda}(\veck, \omega_n) = \mathbf{v}_\veck\Big(\gamma_0(\veck, \omega_n)\tau_0 + i\gamma_1(\veck, \omega_n)\tau_1 + \gamma_3(\veck, \omega_n)\tau_3\Big),
  \label{eqn:currentVertex}
\end{equation}
where explicit expressions for the vertex components $\gamma_n(\veck, \omega_n)$ are given in  Appendix~\ref{appendix_vertex_corrections}.  The final result for the 2D superfluid density is 
\begin{equation}
    \begin{aligned}
    \rho^\mathrm{2D}_s(T) & = \mu_0 {\cal Q}^{xx} = \frac{8 \mu_{0} e^{2} T}{N} \\ 
    &\!\!\!\!\!\!\!\!\!\!\!\!\times \sum_{\omega_n>0}\sum_\veck \left(\mathbf{v}_\veck^j\right)^{2} \frac{\tilde{\Delta}_{\veck,n}^{2} \gamma_{0}(\veck, \omega_n)-\tilde{\omega}_{\veck,n} \tilde{\Delta}_{\veck,n} \gamma_{1}(\veck, \omega_n)}{\left(\tilde\omega_{\veck,n}^2 + \tilde\Delta_{\veck,n}^2 + \xi_{\veck}^2\right)^2}\;.
    \end{aligned}\label{rhos2D}
\end{equation}

The alert reader will note that we have presented formulas in which the $\veck$-sum is performed over the whole Brillouin zone, rather than over the Fermi surface itself as in the previous work of Refs.~\onlinecite{Lee-Hone:2017,Lee-Hone:2018,LeeHone2020}.  This is done both because the kernel in the energy integration of Eq.~(\ref{rhos2D}) does not necessarily fall off rapidly enough near the Fermi surface, and because quantitative inaccuracies arise if calculations are carried out on the Fermi surface near those dopings where the van Hove singularity is close to the Fermi level.  A comparison of the two approaches is given in Appendix~\ref{appendix_van_Hove}, along with details of the implementation of the full-Brillouin-zone sum. 

Sufficiently far from the van Hove singularity, where it is possible to linearize the spectrum $\xi_\veck$ and carry out the $\xi$ integration,  Eq.~(\ref{rhos2D}) can safely be recast as a Fermi surface average.  Including vertex corrections, this leads to the following expression for $\rho^\mathrm{2D}_s$:
\begin{equation}
    \begin{split}
    &\rho^\mathrm{2D}_s(T) = 2 \pi \mu_{0} e^{2} N_0 T \\ 
   & \times \sum_{\omega_n>0} \Bigg\langle  \left(\mathbf{v}_\veck^j\right)^{2} \frac{\tilde{\Delta}_{\veck,n}^{2} \gamma_{0}(\veck, \omega_n)-\tilde{\omega}_{\veck,n} \tilde{\Delta}_{\veck,n} \gamma_{1}(\veck, \omega_n)}{\left(\tilde{\omega}^2_{\veck,n} +\tilde{\Delta}_{\veck,n}^{2}\right)^{3/2}} \Bigg\rangle\FSavg,
    \end{split}
\end{equation}
which reduces to the familiar form for superfluid density used, e.g., in Refs. \onlinecite{Lee-Hone:2017,Lee-Hone:2018,LeeHone2020} in the limit $\gamma_0 \to 1$ and $\gamma_1 \to 0$ when the impurity potentials are pointlike.

\section{Results}

The \textit{ab-initio} calculations of impurity potentials are presented in Appendix~\ref{appendix_impurities}, in Tables~\ref{tab:impuritiesAOV} to \ref{tab:impuritiesTl}.  DFT calculations were initially performed for LSCO, for the undoped supercells La$_{35}$SrCu$_{18}$O$_{72}$ and La$_{36}$Cu$_{18}$O$_{71}$.  In LSCO, the density of states has a particularly strong doping dependence: at a hole doping of around 19\%, the Fermi surface undergoes a Lifshitz transition and a van Hove singularity passes through the Fermi level.  To understand how this influences screening, two further sets of calculations were carried out: one at $p = p_\mathrm{vH}$, with electron count adjusted to place the van Hove singularity at the Fermi level; and an even more overdoped calculation at $p_\mathrm{vH} + 7\%$. Proximity to the van Hove singularity does indeed have a significant effect on the potentials, particularly for the near apical oxygen vacancy V$_\mathrm{O}^\mathrm{near}$.  In Table~\ref{tab:impuritiesAOV}, we see that for undoped La$_{36}$Cu$_{18}$O$_{71}$, the impurity potential for V$_\mathrm{O}^\mathrm{near}$ is rather extended: the on-site energy $V_0$ is comparable in magnitude to the combined energies of the four nearest-neighbor sites ($4 \times V_2$).  This changes markedly at $p_\mathrm{vH}$: V$_\mathrm{O}^\mathrm{near}$ becomes essentially pointlike, with $V_0 \gg V_2, V_4$. As expected, the screening effect diminishes somewhat beyond the van Hove singularity: at $p_\mathrm{vH} + 7\%$ the on-site energy $V_0$ increases by 6\%, while the nearest-neighbour energy $V_2$ doubles, leading to a slightly more extended impurity potential. Proximity to the van Hove singularity also has a systematic effect on the screening of the potential of the Sr dopant, as seen in Table~\ref{tab:impuritiesSr}.  However, the changes are not as strong, nor as important: unlike V$_\mathrm{O}^\mathrm{near}$, the Sr$^\mathrm{near}$ defect is plaquette-centered, so its impurity potential is inherently extended. Results for the Cu defect in Tl-2201 are presented in Table~\ref{tab:impuritiesTl}. Here DFT calculations were only carried out for the undoped Tl$_{35}$Ba$_{36}$Cu$_{19}$O$_{108}$ supercell: there is no Lifshitz transition in Tl-2201 in the relevant doping range and, as discussed in Sec.~\ref{subsec:dopant_potential}, the Cu--Tl substitution already has a hope-doping effect.

As mentioned above, the spatially extended nature of the realistic disorder model gives rise to impurity matrix elements $V_{\veck,\veckP}$ with very strong momentum dependence. This, combined with anisotropic electronic structure, leads to elastic scattering rates that vary strongly around the Fermi surface, something that is a well-established part of cuprate phenomenology.\cite{Hussey:1996eb,Ioffe:1998p386,Valla:2000en,Abrahams:2000hr,Varma:2001bb,Kaminski:2005ge,Plate:2005,AbdelJawad:2006df,Yamasaki:2007hx,Yoshida:2007,Chang:2008cb,Grissonnanche:2021hw}  Recognizing that our calculation
 includes only disorder contributions, and not inelastic
scattering, the normal-state elastic scattering rate $\Gamma_N(\phi)$ is plotted in Figs.~\ref{fig:GammaN_TcTc0}(a) and (b), for LSCO and \mbox{Tl-2201}.   $\Gamma_N(\phi)$ is defined to be the imaginary part of the $\tau_0$ self energy above $T_c$, evaluated on the Fermi surface, at angle $\phi$.  The scattering rate has significant angle dependence in both materials, being strongest in both cases at the antinodes, with minima along the zone diagonals, in qualitative agreement with ARPES\cite{Valla:2000en,Kaminski:2005ge,Plate:2005,Yamasaki:2007hx,Yoshida:2007,Chang:2008cb} and transport\cite{Hussey:1996eb,AbdelJawad:2006df,Grissonnanche:2021hw} experiments on various cuprate materials. In LSCO, the calculated $\Gamma_N(\phi)$ has very little doping dependence, but in \mbox{Tl-2201}, both the magnitude and angle dependence of $\Gamma_N(\phi)$ increase significantly with hole doping.  

ARPES experiments do not, in general, have sufficient resolution to allow a quantitative comparison with our calculated $\Gamma_N(\phi)$.  Clearer insights come from angle-dependent magnetoresistance (ADMR), with $\Gamma(\phi)$ accessible through detailed, Boltzmann-transport fits to the ADMR data. We are not aware of ADMR measurements on LSCO, but results have recently been reported for Nd-doped LSCO,\cite{Grissonnanche:2021hw} in which the Nd dopants shift the Lifshitz transition upwards from $p = 19$\% to \mbox{$p = 24$\%}.  By carrying out these measurements at multiple temperatures, elastic and inelastic contributions to $\Gamma(\phi)$ have been separated. The results from Ref.~\onlinecite{Grissonnanche:2021hw} reveal an elastic scattering rate in Nd-LSCO  qualitatively similar to our calculations, with minima along the nodal directions and maxima at the antinodes. Converting the ADMR values to temperature units, and remembering that the experimentally accessible scattering rate is $2 \Gamma_N(\phi)$, the nodal minimum in \mbox{Nd-LSCO} corresponds to \mbox{$\Gamma_\mathrm{ADMR}^\mathrm{elastic}(\phi\!=\!45^\circ) = 35$~K}. This is somewhat smaller than our calculated nodal minimum for LSCO in Fig.~\ref{fig:GammaN_TcTc0}(a), $\Gamma_N(\phi\!=\!45^\circ) \approx 65$~K, which assumes a concentration of apical oxygen vacancies $n_{\mathrm{V}_\mathrm{O}} = 6\%$.  We have also carried out calculations for $n_{\mathrm{V}_\mathrm{O}} = 4\%$ and find $\Gamma_N(\phi\!=\!45^\circ) \approx 43$~K.  This is significantly closer to the ADMR value, and indicates that a lower concentration of apical oxygen vacancies in the \mbox{Nd-LSCO} crystals could easily account for the difference.  The antinodal scattering rate is another matter, however.  In our calculations, the angular variation in $\Gamma_N$ is slightly less than a factor of 2.  The scattering rate inferred from ADMR varies by a factor 8 around the Fermi surface, reaching an antinodal maximum of 275~K in $\Gamma_N$ units.  We suspect that this discrepancy is an artifact of the Fermi-surface-based Boltzmann-transport approach used to model the ADMR data, which breaks down in the vicinity of a van Hove singularity for the reasons we give in Appendix~\ref{appendix_van_Hove}, requiring full-Brillouin-zone $\mathbf{k}$-sum methods instead.  

For overdoped Tl-2201 ($T_c = 15$~K), the detailed interpretation of ADMR data in Ref.~\onlinecite{AbdelJawad:2006df} found an elastic scattering rate corresponding to \mbox{$\Gamma_N = 20$~K}, with no significant variation around the Fermi surface. This is close to what we calculate for Tl-2201 at optimal doping, but not with the doping dependence we show in Fig.~\ref{fig:GammaN_TcTc0}(b).  A detailed investigation of the behavior of the normal state scattering rate is beyond the scope of the current work,  but it is worth emphasizing two key differences between LSCO and Tl-2201 that must play an important role in any attempt to do so. First, LSCO passes through a Lifshitz transition in the overdoped superconducting range, with the van Hove singularity significantly enhancing its antinodal density of states; the Tl-2201 Fermi surface is far from the vHS, with a relatively isotropic density of states. And second, the elastic scatterers in Tl-2201 are located much further from the CuO$_2$ planes than in LSCO, with concomitantly softer and longer-range potentials, as can be seen in Tables~\ref{tab:impuritiesAOV} to \ref{tab:impuritiesTl}.

For comparison purposes, the values of $\Gamma_N$ used in the point-scattering models of Refs.~\onlinecite{Lee-Hone:2017,Lee-Hone:2018,LeeHone2020} are also shown in Figs.~\ref{fig:GammaN_TcTc0}(a) and (b), and correspond quite closely to the minima of $\Gamma_N(\phi)$ in LSCO and \mbox{Tl-2201}.  This leads to an important insight --- when defects are spatially extended, significantly higher scattering rates can be tolerated, particularly near the antinodes. This is because extended defects are inherently forward scattering. Near the antinodes, forward scattering couples states for which the \mbox{$d$-wave} gap has the same sign, a process akin to Anderson's theorem,\cite{Anderson:1959p2774} and harmless to superconductivity.\cite{Kee:2001bw,Graser:2007kp}  This can further be seen in Figs.~\ref{fig:GammaN_TcTc0}(c) and (d), where plots of clean-limit transition temperature, $T_{c0}(p)$, show the pairing strength needed to produce the parabolic ``superconducting domes" of LSCO and Tl-2201. We see that the required pairing strength is somewhat lower than for the point scatterers of Refs.~\onlinecite{Lee-Hone:2017,Lee-Hone:2018,LeeHone2020}, despite the larger average scattering rate in Figs.~\ref{fig:GammaN_TcTc0}(a) and (b). For LSCO this is particularly welcome, as the large values of $T_{c0}(p)$ needed in Refs.~\onlinecite{Lee-Hone:2017,Lee-Hone:2018,LeeHone2020} were a distinct weakness of the point-scattering approach. 
It is intriguing that this suggests an intrinsic pairing scale of about 80~K in doped La$_2$CuO$_4$, slightly higher than the 70~K of superoverdoped Ba$_3$Cu$_2$O$_{4-\delta}$.\cite{Li2019}  As expected, in the Tl-2201 system, the $T_c$ suppression by disorder is considerably weaker than in LSCO.

\begin{figure}[t]
    \centering
        \includegraphics[width=0.8\columnwidth,scale=1.0]{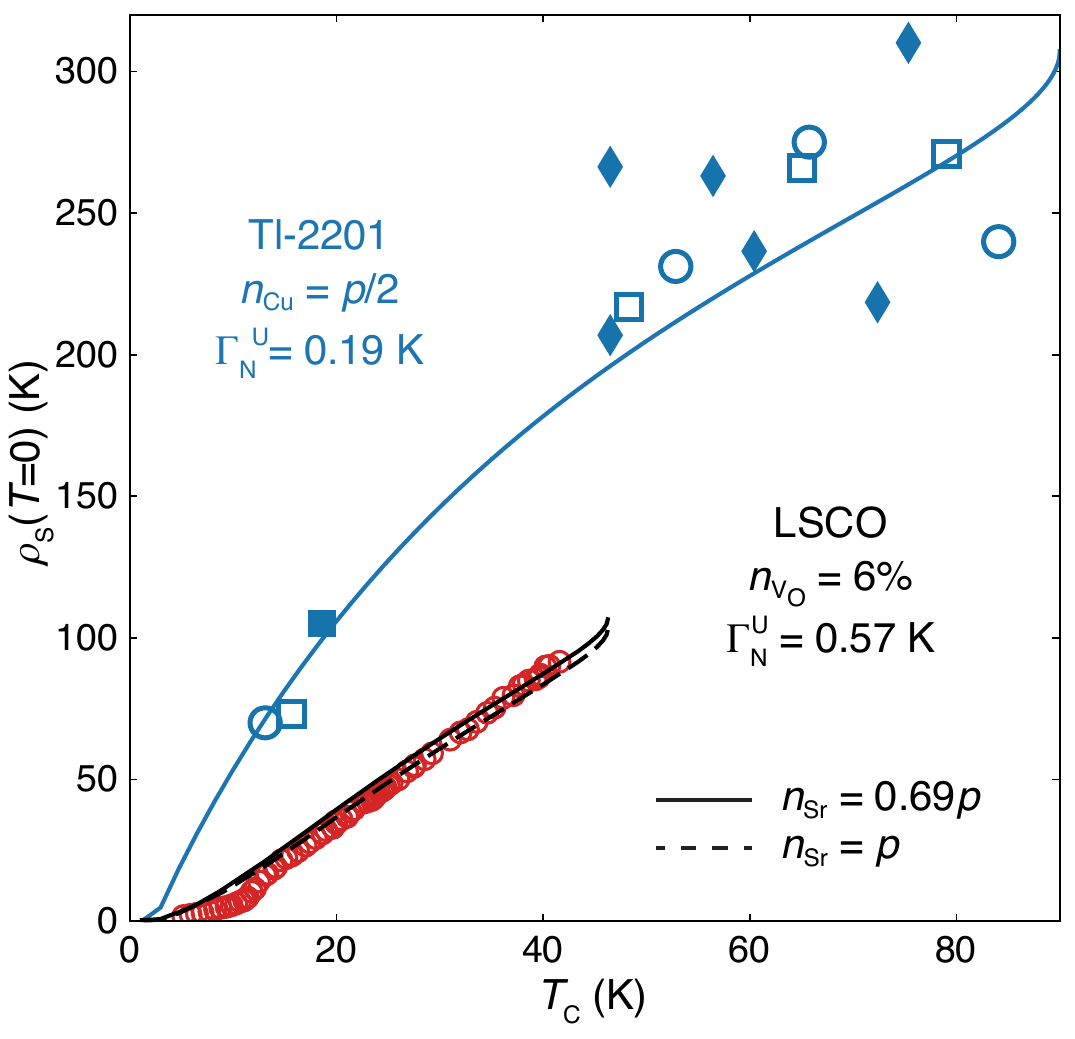}
    \caption{Zero-temperature superfluid density as a function of transition temperature $T_c$ for LSCO and Tl-2201. Disorder parameters are as shown including, for LSCO, two different doping dependences of Sr concentration, $n_\mathrm{Sr}(p)$. LSCO data: {\LSCOColor \CircPipe} MBE thin-film mutual inductance. \cite{Bozovic:2016ei}  Tl-2201 data: $\TlColor\blacksquare$ single-crystal microwave;\cite{Deepwell:2013uu} \mbox{$\TlColor\blacklozenge$ single-crystal $\mu$SR;\cite{Brewer2015}} {\TlColor\CircPipe}, {\TlColor\SquarePipe} polycrystalline $\mu$SR. \cite{Niedermayer1993,Uemura1993}}
    \label{fig:rho0VTc}
\end{figure}

\begin{figure*}[t]
    \centering
        \includegraphics[width=
        0.95\linewidth,scale=1.0]{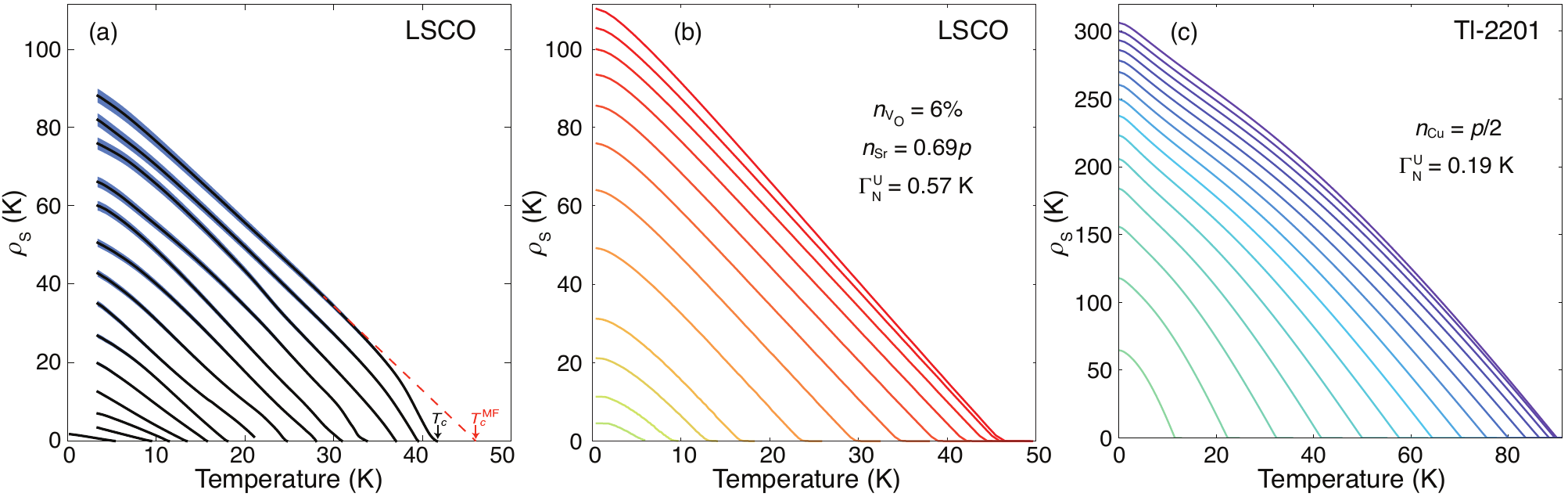}
    \caption{Temperature-dependent superfluid density of overdoped LSCO and Tl-2201. (a) Superfluid density data for LSCO from Ref.~\onlinecite{Bozovic:2016ei}, with shading indicating the stated 2\% experimental uncertainty. The dashed line indicates the construction used to estimate the corresponding mean-field transition temperature $T_c^\mathrm{MF}$. (b,c) Superfluid density for LSCO and Tl-2201 calculated in the realistic disorder model, for the disorder parameters shown, as doping (and therefore $T_c$) are varied.}
    \label{fig:superfluid_vs_T}
\end{figure*}

Figure~\ref{fig:rho0VTc} shows zero-temperature superfluid density as a function of transition temperature $T_c$ in LSCO and Tl-2201. The \textit{ab-initio} calculations do a remarkably good job of capturing the experimental behavior, particularly in the case of Tl-2201, where the sole assumption is that the concentration of out-of-plane scatterers is given by $n_\mathrm{Cu} = p/2$.  In LSCO, the data of Ref.~\onlinecite{Bozovic:2016ei} are well described by a calculation based on 6\% apical oxygen vacancies. In Appendix~\ref{appendix_strontium}, the effect of varying the concentration of Sr and V$_\mathrm{O}$ defects in LSCO is explored in more detail.  The pointlike potential of the apical oxygen vacancies has a significant pair-breaking effect; the spatially extended potential of the Sr defects renders them fairly benign to the superfluid density, as can be seen in Fig.~\ref{fig:rho0VTc} by comparing the traces for $n_\mathrm{Sr} = p$ and $n_\mathrm{Sr} = 0.69 p$.

Figure~\ref{fig:superfluid_vs_T} shows the temperature dependence of $\rho_s$ for the same \textit{ab-initio} potentials as in Fig.~\ref{fig:rho0VTc}, along with the experimental data for LSCO from Ref.~\onlinecite{Bozovic:2016ei}.  As expected from Refs.~\onlinecite{Lee-Hone:2017,LeeHone2020}, the temperature dependence is remarkably linear down to temperatures of order 1~K, at which point the materials enter a gapless regime indicative of a narrow impurity band, controlled in our model by the presence of a small density of unitarity scatterers.  In general, the agreement with the data of Ref.~\onlinecite{Bozovic:2016ei} shown in Fig.~\ref{fig:superfluid_vs_T}(a) is excellent over the entire temperature and doping range, with the exception of deviations near $T_c$, which are discussed in Ref.~\onlinecite{Lee-Hone:2017}; and for the very lowest $T_c$ sample, which was discussed in detail in Ref.~\onlinecite{Bozovic:2018dx}.  We do not currently understand the latter discrepancy: the disorder-averaged theory suggests that significant curvature will occur in $\rho_s(T)$ at the lowest temperatures, but this is not visible in the experimental data on that sample.  It may be that the very low $T_c$ samples are more inhomogeneous and have such low superfluid density that they are indeed dominated by phase fluctuations rather than Bogoliubov quasiparticles.
        
\section{Discussion}
\label{sec:discussion}

\subsection{Critical assessment of approach}

Here we present what we regard as the strong and weak points of the analysis we have presented. A major advantage of our approach, a simple extension of ``dirty $d$-wave theory", is certainly its simplicity:   one can work in the superconducting state with a theory of well-defined quasiparticles and calculate essentially any observables.  We have now updated this theory to start from \textit{ab-initio} calculations of the relevant impurity strengths in the systems of interest, and shown that the results continue to agree well with recent superfluid density experiments;  the number of free parameters not precisely constrained by experimental data or the \textit{ab-initio} calculations has thus been significantly reduced.  Furthermore, we show in Appendix~\ref{appendix_van_Hove} that the van Hove singularity, which naively could give rise to anomalies in such a theory at dopings when it approaches the Fermi level, is in fact rather benign when properly treated using momentum-sum methods rather than Fermi-surface averages.

Nevertheless, the present approach could be improved in several ways.  First, as we have noted above, one could study the out-of-plane dopant impurities within the $t$-matrix rather than Born approximation.  This would be more accurate, but makes calculation of the corresponding vertex corrections to two-particle properties significantly more expensive.  We have argued that, for the calculated impurity potential values, \mbox{$t$-matrix} and Born approximation calculations must give essentially identical results.

A complete theory would also account explicitly for the 
doping effect of Sr/La,  Cu/Tl, and O vacancy impurities on the electronic structure.  This approach, i.e., assuming that every impurity dopes the system, was adopted, e.g., in Ref.~\onlinecite{Li2021}.  While appealing in its simplicitly, it does not reflect the complexity of different impurities and dopants in the cuprates.  We have therefore chosen to calculate the \textit{ab-initio} scattering potentials directly, and have incorporated the changes in electronic structure by adopting the ARPES-derived tight binding models interpolated across the doping range.   A direct implementation of \textit{ab-initio} doping is beyond the scope of the present theory, but it might be feasible in some future implementation of DFT+DMFT or similar realistic theories of electronic structure incorporating correlations. 

In addition, we have performed our calculations of superconducting-state observables on a two-dimensional lattice of Cu $3d_{x^2 - y^2}$ orbitals.  We thereby miss explicit inclusion of O and other out-of-plane ionic degrees of freedom in the tight-binding  electronic structure, except insofar as they contribute to the single hybridized  Cu--O band near the Fermi level. In some cuprates, 3D effects can be important at the level of electronic structure, particularly in tight-binding descriptions of the CuO$_2$ plane near van Hove singularities, as shown in Ref.~\onlinecite{Fresard2019}.

Finally, we note that  as in previous work, we have assumed that the pairing interaction binding electrons into a $d$-wave state can be adequately described in the weak-coupling limit, by a separable pairing interaction, the strength of which is parameterized by an assumed $T_{c0}(p)$, chosen to reproduce the experimental superconducting phase diagram. In a complete theory, the weakening of the clean-limit pairing interaction with increasing doping in the overdoped regime should be calculated from microscopics as well. 
For example, one might expect the doping and disorder to modify the particle--hole susceptibility as one moves across the phase diagram, thereby modifying the underlying spin-fluctuation pairing strength.\cite{Scalapino_Maier2017}  Including simple models of the disorder self-energy in such theories has already been attempted.\cite{Maier2020}  It will certainly be worthwhile to explore combining these approaches with our microscopic, material-specific disorder models. 

 \subsection{Electronic inhomogeneity}

In this paper, we have explored the predictions of a simple theory describing a homogeneously disordered system, i.e., we have assumed that defects are distributed randomly and perturb the otherwise homogeneous system only weakly.  This allows the application of a conventional disorder-averaged ``effective medium" approach, which can be considered a simple generalization of the successful ``dirty $d$-wave" theory applied earlier to cuprates.  In such a picture, the temperature dependence of thermodynamic and transport properties in the superconducting state is dominated by Bogoliubov quasiparticles.  Furthermore, the disappearance of superconductivity on the overdoped side of the dome must be considered to be a consequence of a combination of a weakening of the $d$-wave pairing interaction and disorder (see, e.g., Ref.~\onlinecite{Maier2020}).  

On the other hand, there are several studies of cuprates generally, and of LSCO in particular, which propose a very different picture of the overdoped systems,
one in which larger-scale inhomogeneity is {\it intrinsic} to the materials, and therefore $T_c$ is determined not by uniform suppression of the gap, but by the destruction of the Josephson coupling between grains by phase fluctuations.  For example, a recent study\cite{Tranquada2022} considering several measures of superconductivity on an overdoped LSCO crystal found strong nanoscale inhomogeneity in STM and an onset of diamagnetism many degrees above the onset of zero resistance, consistent with a theory of granular superconductivity by Spivak et al.\cite{Spivak2008}    There is evidence from STM on many Bi-based cuprates of substantial inhomogeneity of local gaps and other properties of these samples.\cite{Fujita2011,Tromp2022}

Nevertheless, thermodynamic measurements tend to infer much narrower spatial distributions of superconducting properties in the overdoped regions than STM, particularly in the cleaner YBCO material\cite{Loram2004,Bobroff2004} (overdoped with Ca).  Even in LSCO, sample quality varies dramatically between crystals and thin films of various types, and there is to our knowledge no evidence for gross inhomogeneity in the epitaxially grown films of Ref.~\onlinecite{Bozovic:2016ei} with which we compare here.   Recently, a careful study of Ca substitution concluded that reducing the disorder strength can dramatically extend the $T_c$ dome of LSCO,\cite{Kim2021} supporting the notion that the destruction of superconducitivity depends strongly on the scattering potentials of the dopants, as assumed here.
From the theoretical standpoint, while it has been proposed that highly disordered $d$-wave superconductors can exhibit responses consistent with a granular picture \cite{Li2021} a recent study searching for this effect found no significant self-organization of regions of well-defined phase.\cite{Breio2022}

It therefore seems likely that, while granular, phase-fluctuation-dominated behavior may indeed exist in some overdoped samples, it is not {\it intrinsic} to cuprates, but rather due to chemical inhomogeneities that arise naturally in the growth process of some samples of some materials.  In this case, the description of the somewhat idealized disorder-averaged dirty $d$-wave theory may indeed prove adequate for most of the overdoped phase diagram, until the falling superfluid density induces strong phase fluctuations\cite{Emery1995} that may overcome the quasiparticle responses  for some observables in samples  with very small $T_c$.  Furthermore, it seems to us unlikely that a theory of weakly Josephson-coupled grains can explain the superfluid density, optical conductivity, specific heat and thermal conductivity at the same semiquantitative level put forward { here and} in Refs.~\onlinecite{Lee-Hone:2017,Lee-Hone:2018,LeeHone2020}.

\section{Conclusions}
 As discussed in the introduction, we have started from the point of view that the experimental data on the overdoped side of the cuprate phase diagram can be explained within the Landau-BCS paradigm, provided the details of the disorder present as a result of the doping process are properly accounted for, with appropriate Fermi liquid renormalizations of quasiparticle masses fixed by experiment.  Within this scheme, the dopant disorder is taken to be  weak enough to apply the Born approximation for impurity scattering. A small concentration of strong, pointlike scatterers, representing in-plane defects such as Cu vacancies, has also been included, in order to reproduce the small gapless impurity band observed at the lowest temperatures.   The impurity potentials themselves have now been calculated from first principles, scaled only by an overall bandwidth renormalization, yielding a tractable theory within which any superconducting state observable  can be calculated.   We have shown here that for the two cuprates that can be easily overdoped, these simple conventional procedures produce results consistent with the ``anomalous" behavior of the superfluid density that has been observed.
 
\begin{acknowledgments}
We are grateful for useful discussions with J.~S.~Dodge, S.~A.~Kivelson, T.~A.~Maier, D.~J.~Scalapino, J.~E.~Sonier, and J.M. Tranquada.  D.M.B.\  acknowledges financial support from the Natural Science and Engineering Research Council of Canada.  P.J.H.\ acknowledges support from NSF-DMR-1849751.
V.M.\ was supported by NSFC \mbox{Grant No.~11674278} and by the priority program of the Chinese Academy of Sciences \mbox{Grant No.~XDB28000000}. The first-principles calculations in this work (X.K. and T.B.) were conducted at the Center for Nanophase Materials Sciences, a US Department of Energy Office of Science User Facility, operated at Oak Ridge National Laboratory. We used resources of the Compute and Data Environment for Science (CADES) at the Oak Ridge National Laboratory, which is supported by the Office of Science of the U.S.\ Department of Energy under Contract No.\ DE-AC05-00OR22725. In addition we used resources of the National Energy Research Scientific Computing Center, a DOE Office of Science User Facility supported by the Office of Science of the U.S.\ DOE under Contract No.\ DE-AC02-05CH11231.
\end{acknowledgments}

\begin{figure}[t]
    \centering
        \includegraphics[width=\linewidth,scale=1.0]{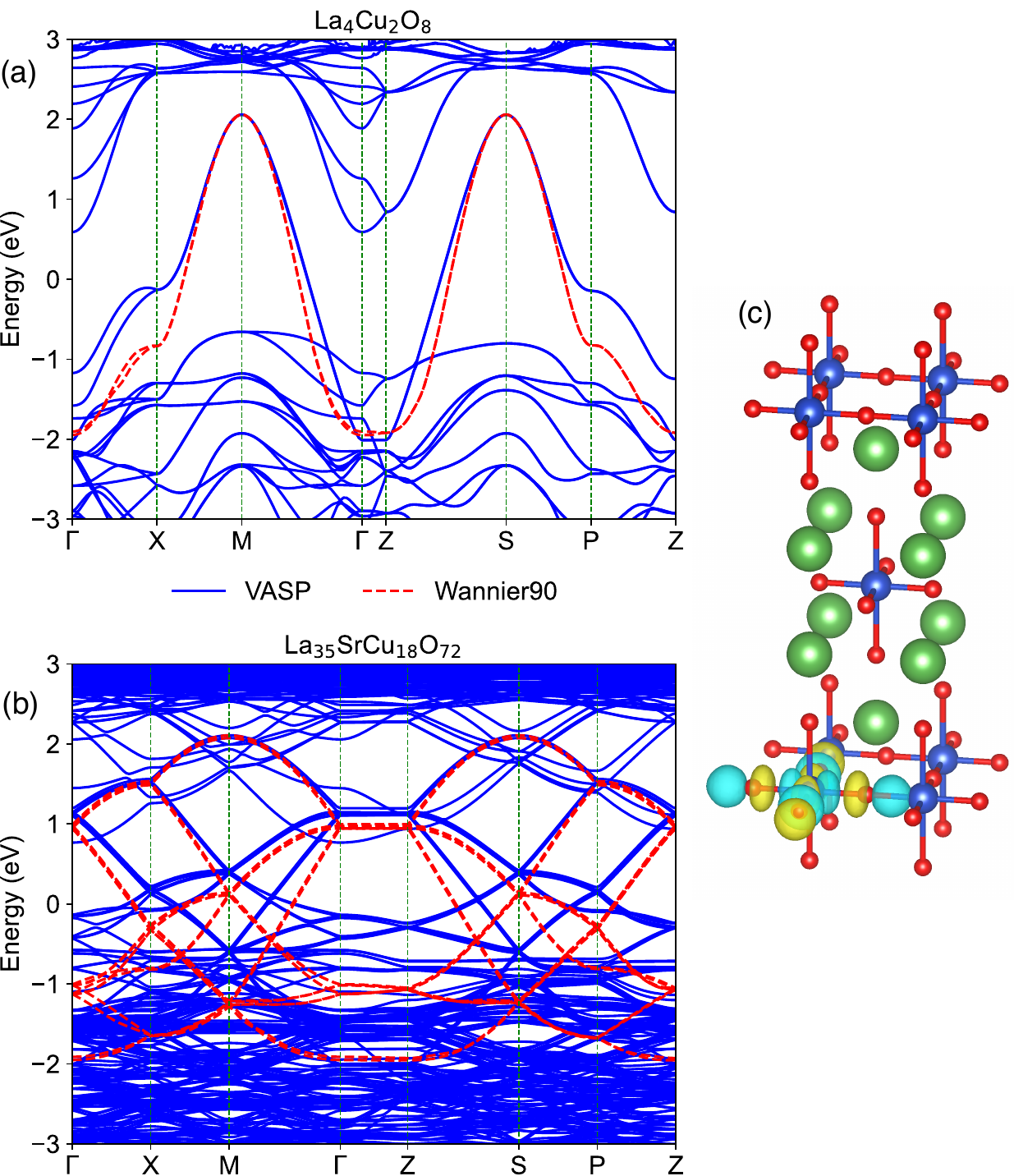}
    \caption{Comparison of the VASP DFT bandstructure (solid lines) with the Wannierized bands (dashed lines) used to generate the tight-binding impurity potentials, for (a) the undoped La$_4$Cu$_2$O$_8$ normal cell and (b) the La$_{35}$SrCu$_{18}$O$_{72}$ supercell, containing one Sr dopant. The high symmetry points are given by: 
    \mbox{$\Gamma = (0,0,0)$},
    \mbox{$X = (\tfrac{1}{2},0,0)$}, 
    \mbox{$M = (\tfrac{1}{2},\tfrac{1}{2},0)$}, 
    \mbox{$Z = (0,0,\tfrac{1}{2})$}, 
    \mbox{$S = (\tfrac{1}{2},\tfrac{1}{2},\tfrac{1}{2})$},
    \mbox{$P = (\tfrac{1}{2},0,\tfrac{1}{2})$}, and 
    \mbox{$Z = (0,0,\tfrac{1}{2})$}.
    For the La$_4$Cu$_2$O$_8$ and La$_{35}$SrCu$_{18}$O$_{72}$ bands these are expressed in the reciprocal basis vectors of the orthogonal La$_4$Cu$_2$O$_8$ and La$_{35}$SrCu$_{18}$O$_{72}$ unit cells, respectively. (c) The Wannier function, displaying well-localized \mbox{Cu $d_{x^2 - y^2}/$O $p_{x,y}$} anti-bonding character. 
    We emphasize that the Wannier bands are only used in the determination of the impurity potentials: the empirical ARPES band structure is the basis for all other calculations.} 
    \label{fig:band_comparison}
\end{figure} 

\begin{table*}[t]
\renewcommand{\arraystretch}{1.7}
\begin{tabular}
{|c|c|c|c|c|c|c|c|c|}
\hline
\multirowcell{3}{Defect\\\&\\location} & \multirowcell{3}{Impurity\\term} & \multicolumn{2}{c|}{\multirowcell{2}{Representative\\coordinates}} & \multicolumn{3}{c|}{\multirowcell{2}{Impurity potential \\ $\Delta E/|t| \times 1000\big.$}} & \multirowcell{3}{Form factor\\$f(\mathbf{q})\Big.$ or $f(\mathbf{k},\mathbf{k}^\prime)$}& \multirowcell{3}{Form factor\\$f(\mathbf{q} = 0\big.) \equiv f(\mathbf{k} = \mathbf{k}^\prime)\Big.$}\\
&  & \multicolumn{2}{c|}{} & \multicolumn{3}{c|}{} & & \\
\cline{3-7}
  &  & $\;\;\;\,\mathbf{R}/a\;\;\;\,$ & $\mathbf{R}^\prime/a$ & undoped & \;\;\;\;$p_\mathrm{vH}$\;\;\;\;& $p_\mathrm{vH}\!+\!7\%$ & & \\ 
  
\hline\hline

\multirowcell{7}{V$_\mathrm{O}^\mathrm{near}$\\$z = 2.4$~\AA\\site-\\centered} 
& $V_0 = V_\mathbf{R}$ & (0,0) & -- & -323.96 & -358.68 & -381.55 & 1 & 1\\ 
\cline{2-9}
 & $V_2 = V_\mathbf{R}$ & (1,0) & -- & -104.36 & -12.90 & -25.40 & $2\big[\cos(q_x) + \cos(q_y)\big]$ & 4\\
\cline{2-9}
 & $V_4 = V_\mathbf{R}$ & (1,1) & -- & -39.01 & 11.09 & -13.77 & $4\;\cos(q_x)\cos(q_y)$ & 4\\
\cline{2-9}
 &  $\delta t_0 = \delta t_{\mathbf{R},\mathbf{R}^\prime}$ & (0,0) & (1,0) & 8.08 & 14.83 & 16.52 & $2\big[\cos(k_x)+\cos(k_y)+\cos(k_x')+\cos(k_y')\big]$ & $4\big[\cos(k_x)+\cos(k_y)\big]$\\
\cline{2-9}
 &  \multirowcell{2}{$\delta t_2 = \delta t_{\mathbf{R},\mathbf{R}^\prime}$}& \multirowcell{2}{(1,0)} & \multirowcell{2}{(1,1)} & \multirowcell{2}{4.97} & \multirowcell{2}{8.96} & \multirowcell{2}{9.64} & \multirowcell{2}{4$\big(\cos(k_x - k_x')\big[\cos(k_y)+\cos(k_y')\big]\Big.$\\$+\;\cos(k_y - k_y')\big[\cos(k_x)+\cos(k_x')\big]\big)$}& \multirowcell{2}{$8\big[\cos(k_x)+\cos(k_y)\big]$}\\
 & & & & & & & &\\
\cline{2-9}
 &  $\delta t_0^\prime = \delta t_{\mathbf{R},\mathbf{R}^\prime}$ & (0,0) & (1,1) & 4.36 & 3.20 & 4.84 & $4\big[\cos(k_x)\cos(k_y) + \cos(k_x')\cos(k_y')\big]$ & $8 \cos(k_x)\cos(k_y)$\\
 
\hline\hline

\multirowcell{9}{V$_\mathrm{O}^\mathrm{far}$\\$z = 4.2$~\AA\\plaquette-\\centered} 
& $V_1 = V_\mathbf{R}$ & $\left(\frac{1}{2},\frac{1}{2}\right)$ & -- & -52.48 & -45.78 & -59.18 & $4\cos\left(\frac{q_x}{2}\right)\cos\left(\frac{q_y}{2}\right)$ & 4\\ 
\cline{2-9}
 & $V_3 = V_\mathbf{R}$ & $\left(\frac{1}{2},\frac{3}{2}\right)$ & -- & -13.59 & -7.72 & -17.31 & $4\Big[\cos\Big(\frac{q_x}{2}\Big)\cos\Big(\frac{3q_y}{2}\Big)+ \cos\Big(\frac{3 q_x}{2}\Big)\cos\Big(\frac{q_y}{2}\Big)\!\Big]$ & 8\\
\cline{2-9}
  &  \multirowcell{2}{$\delta t_1 = \delta t_{\mathbf{R},\mathbf{R}^\prime}$}& \multirowcell{2}{$\left(\frac{1}{2},\frac{1}{2}\right)$} & \multirowcell{2}{$\left(\frac{1}{2},-\frac{1}{2}\right)$} & \multirowcell{2}{12.29} & \multirowcell{2}{18.68} & \multirowcell{2}{20.09} & \multirowcell{2}{$8\Big[\cos\Big(\frac{k_x + k_y}{2}\Big)\cos\Big(\frac{k_x' - k_y'}{2}\Big)$\\$ + \cos\Big(\frac{k_x - k_y}{2}\Big)\cos\Big(\frac{k_x' + k_y'}{2}\Big)\Big]$} &  \multirowcell{2}{$8\big[\cos(k_x)+\cos(k_y)\big]$}\\
 & & & & & & & &\\
\cline{2-9}
  &  \multirowcell{4}{$\delta t_3 = \delta t_{\mathbf{R},\mathbf{R}^\prime}$} & \multirowcell{4}{$\left(\frac{1}{2},\frac{1}{2}\right)$} & \multirowcell{4}{$\left(\frac{1}{2},\frac{3}{2}\right)$} & \multirowcell{4}{5.95} & \multirowcell{4}{5.85} & \multirowcell{4}{6.21} & \multirowcell{4}{$4\Big[\cos\Big(\frac{2 k_x + 2 k_y - k_x' - k_y'}{2}\Big)\cos\Big(\frac{k_x - k_y}{2}\Big)$\\
  $+ \cos\Big(\frac{2 k_x - 2 k_y - k_x' + k_y'}{2}\Big) \cos\Big(\frac{k_x + k_y}{2}\Big)$\\
$  + \cos\Big(\frac{k_x + k_y - 2 k_x' - 2 k_y' }{2}\Big)  
  \cos\Big(\frac{k_x' - k_y'}{2}\Big)$\\
$  + \cos\Big(\frac{k_x - k_y -2k_x' + 2 k_y' }{2}\Big) 
  \cos\Big(\frac{k_x' + k_y'}{2}\Big)\Big]$}&  \multirowcell{4}{$8\big[\cos(k_x)+\cos(k_y)\big]$}\\
 & & & & & & & &\\
 & & & & & & & &\\
 & & & & & & & &\\
\cline{2-9}
 &  $\delta t_1^\prime = \delta t_{\mathbf{R},\mathbf{R}^\prime}$ & $\left(\frac{1}{2},\frac{1}{2}\right)$ & $\left(-\frac{1}{2},-\frac{1}{2}\right)$ & 3.66 & 1.01 & 0.51 & $8 \cos\Big(\frac{k_x + k_x' }{2}\Big) \cos\Big(\frac{k_y + k_y'}{2}\Big)$& $8 \cos(k_x)\cos(k_y)$\\
 
\hline
\end{tabular}
\caption{\label{tab:impuritiesAOV} Impurity parameters for the apical oxygen vacancy, V$_\mathrm{O}$, in LSCO. \textit{Ab-initio} potentials from Wannier-projected DFT are tabulated in units of the DFT-derived nearest-neighbor hopping, $|t| = 524.7$~meV, at three hole doping levels: the undoped La$_{36}$Cu$_{18}$O$_{71}$ supercell; $p_\mathrm{vH}$, with the van Hove singularity tuned to the Fermi level; and $p_\mathrm{vH} + 7\%$. Representative values of the coordinates $\mathbf{R}$ and $\mathbf{R}^\prime$ are given in a coordinate system where the defect is located a distance $z$ above the origin. For V$_\mathrm{O}^\mathrm{near}$ ($z = 2.4$~\AA), the origin is centered on a Cu site. For V$_\mathrm{O}^\mathrm{far}$ ($z = 4.2$~\AA), the origin is centered on a CuO$_2$ plaquette. Additional, symmetry-related copies are implied, and when included produce the momentum-dependent form factors shown, with momenta in units of $1/a$, the inverse lattice parameter.  For the site-energy terms, the form factors depend only on the momentum transfer $\mathbf{q} = \mathbf{k} - \mathbf{k}^\prime.$ Form factors for the hopping modifications depend on $\mathbf{k}$ and $\mathbf{k}^\prime$ separately.}
\end{table*}

\appendix

\section{Impurity potentials}    
\label{appendix_impurities}

As summarized above, the calculation of the impurity potential associated with a given defect requires two DFT calculations to be carried out: one for a supercell containing a single impurity; and a reference calculation of the corresponding pure system. For these calculations we employed the Vienna \textit{ab-initio} simulation package (VASP) \citep{Kresse1999} with the generalized gradient approximation of Perdew, Burke and Ernzerhof,\citep{PBE1996} and a plane-wave energy cut-off of $650$ eV. For the reference systems we used the conventional orthogonal La$_4$Cu$_2$O$_8$ and Tl$_4$Ba$_4$Cu$_2$O$_{12}$ cells with with structural parameters corresponding to the Inorganic Crystal Structure Database (ICSD) entries 41643 and 65326 respectively.\cite{icsd} The $k$-grids were taken to be $21 \times 21 \times 6$ and $7 \times 7 \times 6$ for the La$_4$Cu$_2$O$_8$ normal cells and supercells, respectively. For the Tl$_4$Ba$_4$Cu$_2$O$_{12}$ normal cells and supercells, the $k$-grids were taken to be $12 \times 12 \times 2$ and $4 \times 4 \times 2$, respectively. To derive the impurity potentials for the (La,Sr) substitution and the O vacancy impurity potential in LSCO with the additional hole dopings we proceeded as follows. We adjusted the VASP parameter $\mathtt{NELECT}$ in such a way that the van Hove singularity in the supercell lies at the Fermi energy. For the La$_{35}$SrCu$_{18}$O$_{72}$ supercell this required us to subtract 0.11 electrons per Cu from the default value of $\mathtt{NELECT}$. For the La$_{36}$Cu$_{18}$O$_{71}$ supercell this number was 0.26 electrons per Cu instead. This difference can be roughly understood from the fact that a Sr dopes 1 hole, and an O vacancy removes 2 holes: $0.11+1/18\approx 0.26-2/18$. We note that for these impurity potential derivations, $\mathtt{NELECT}$ was adjusted both in the single impurity supercell, and the reference system (La$_4$Cu$_2$O$_8$ in this case) by the same number of  holes per Cu.  Adjusting the NELECT parameter adds a constant positive/negative background potential that essentially mimics the many positively/negatively charged Sr and O vacancy defects that are distributed throughout the sample, \textit{not} the single, isolated impurity for which the \textit{ab-initio} potential is being calculated.  In the single-impurity approach taken in our study, these background corrections are the most straightforward approximation to take into account the potentials induced by the nonzero density of dopants and vacancies present in sample.  In addition to the van Hove hole doping $p_\mathrm{vH}$, we also derived the impurity potentials of the (La,Sr) substitution and the O vacancy in LSCO for the hole doping $p_\mathrm{vH}+7\%$ in which 0.07 extra holes were doped per Cu. In total eleven DFT calculations were performed to derive the impurity potentials of the (La,Sr) substitution and the O vacancy in LSCO for various hole dopings. For the undoped impurity potentials we simulated La$_4$Cu$_2$O$_8$,  La$_{35}$SrCu$_{18}$O$_{72}$ and La$_{36}$Cu$_{18}$O$_{71}$ with the default value of $\mathtt{NELECT}$. For the (La,Sr) potential at $p_\mathrm{vH}$ and $p_\mathrm{vH}+7\%$ we simulated La$_4$Cu$_2$O$_8$ and La$_{35}$SrCu$_{18}$O$_{72}$  in which 0.11 and 0.18 electrons were removed from the default value of $\mathtt{NELECT}$. For the O vacancy potential at $p_\mathrm{vH}$ and $p_\mathrm{vH}+7\%$ we simulated La$_4$Cu$_2$O$_8$ and La$_{35}$SrCu$_{18}$O$_{72}$  in which 0.26 and 0.33 electrons we removed from the default value of $\mathtt{NELECT}$. We emphasize again that in addition to the explicit removal of electrons, the impurities themselves also dope holes and electrons: a Sr dopes 1 hole, and an O vacancy dopes 2 electrons.

\begin{table*}[t]
\renewcommand{\arraystretch}{1.7}
\begin{tabular}
{|c|c|c|c|c|c|c|c|c|}
\hline
\multirowcell{3}{Defect\\\&\\location} & \multirowcell{3}{Impurity\\term} & \multicolumn{2}{c|}{\multirowcell{2}{Representative\\coordinates}} & \multicolumn{3}{c|}{\multirowcell{2}{Impurity potential \\ $\Delta E/|t| \times 1000$}} & \multirowcell{3}{Form factor\\$f(\mathbf{q})$ or $f(\mathbf{k},\mathbf{k}^\prime)$}&\multirowcell{3}{Form factor\\$f(\mathbf{q} = 0\big.) \equiv f(\mathbf{k} = \mathbf{k}^\prime)\Big.$}\\
&  & \multicolumn{2}{c|}{} & \multicolumn{3}{c|}{} & & \\
\cline{3-7}
  &  & $\;\;\;\,\mathbf{R}/a\;\;\;\,$ & $\mathbf{R}^\prime/a$ & undoped & \;\;\;\;$p_\mathrm{vH}$\;\;\;\;  & $p_\mathrm{vH} + 7\%$ &  &\\ 

\hline\hline

\multirowcell{9}{Sr$^\mathrm{near}$\\$z = 1.8$~\AA\\plaquette-\\centered} 
& $V_1 = V_\mathbf{R}$ & $\left(\frac{1}{2},\frac{1}{2}\right)$ & -- & 97.56 & 70.80 & 69.44 & $4\cos\left(\frac{q_x}{2}\right)\cos\left(\frac{q_y}{2}\right)$& 4 \\ 
\cline{2-9}
 & $V_3 = V_\mathbf{R}$ & $\left(\frac{1}{2},\frac{3}{2}\right)$ & -- & 18.28 & 6.99 & 7.73 & $4\Big[\cos\Big(\frac{q_x}{2}\Big)\cos\Big(\frac{3q_y}{2}\Big)+ \cos\Big(\frac{3 q_x}{2}\Big)\cos\Big(\frac{q_y}{2}\Big)\!\Big]$ & 8\\
\cline{2-9}
  &  \multirowcell{2}{$\delta t_1 = \delta t_{\mathbf{R},\mathbf{R}^\prime}$}& \multirowcell{2}{$\left(\frac{1}{2},\frac{1}{2}\right)$} & \multirowcell{2}{$\left(\frac{1}{2},-\frac{1}{2}\right)$} & \multirowcell{2}{-27.84} & \multirowcell{2}{-29.81} & \multirowcell{2}{-30.18} & \multirowcell{2}{$8\Big[\cos\Big(\frac{k_x + k_y}{2}\Big)\cos\Big(\frac{k_x' - k_y'}{2}\Big)$\\ 
  $+ \cos\Big(\frac{k_x - k_y}{2}\Big)\cos\Big(\frac{k_x' + k_y'}{2}\Big)\Big]$} &  \multirowcell{2}{$8\big[\cos(k_x)+\cos(k_y)\big]$}\\
 & & & & & & & &\\
\cline{2-9}
  &  \multirowcell{4}{$\delta t_3 = \delta t_{\mathbf{R},\mathbf{R}^\prime}$}& \multirowcell{4}{$\left(\frac{1}{2},\frac{1}{2}\right)$} & \multirowcell{4}{$\left(\frac{1}{2},\frac{3}{2}\right)$} & \multirowcell{4}{-18.14} & \multirowcell{4}{-15.79} & \multirowcell{4}{-14.21} &\multirowcell{4}{$4\Big[\cos\Big(\frac{2 k_x + 2 k_y - k_x' - k_y'}{2}\Big) 
\cos\Big(\frac{k_x - k_y}{2}\Big)$\\
 $ + \cos\Big(\frac{2 k_x - 2 k_y - k_x' + k_y'}{2}\Big) \cos\Big(\frac{k_x + k_y}{2}\Big)$\\
 $ + \cos\Big(\frac{k_x + k_y - 2 k_x' - 2 k_y' }{2}\Big)  
  \cos\Big(\frac{k_x' - k_y'}{2}\Big)$\\
 $ + \cos\Big(\frac{k_x - k_y -2k_x' + 2 k_y' }{2}\Big) 
  \cos\Big(\frac{k_x' + k_y'}{2}\Big)\Big]$} & \multirowcell{4}{$8\big[\cos(k_x)+\cos(k_y)\big]$}\\
 & & & & & & & &\\
 & & & & & & & &\\
 & & & & & & & &\\
\cline{2-9}
 &  $\delta t_1^\prime = \delta t_{\mathbf{R},\mathbf{R}^\prime}$ & $\left(\frac{1}{2},\frac{1}{2}\right)$ & $\left(-\frac{1}{2},-\frac{1}{2}\right)$ & -5.77 & -5.78 & -5.40 & 
 $8 \cos\Big(\frac{k_x + k_x' }{2}\Big) \cos\Big(\frac{k_y + k_y'}{2}\Big)$ & $8 \cos(k_x)\cos(k_y)$\\

\hline\hline

\multirowcell{7}{Sr$^\mathrm{far}$\\$z = 4.8$~\AA\\site-\\centered} 
& $V_0 = V_\mathbf{R}$ & (0,0) & -- & 30.83 & -9.12 & -22.24 & 1 & 1\\ 
\cline{2-9}
 & $V_2 = V_\mathbf{R}$ & (1,0) & -- & 15.71 & -4.57 & -3.86 & $2\big[\cos(q_x) + \cos(q_y)\big]$ & 4\\
\cline{2-9}
 & $V_4 = V_\mathbf{R}$ & (1,1) & -- & 15.79 & 0.62 & 3.33 & $4\;\cos(q_x)\cos(q_y)$ & 4\\
\cline{2-9}
 &  $\delta t_0 = \delta t_{\mathbf{R},\mathbf{R}^\prime}$ & (0,0) & (1,0) & -10.03 & -9.81 & -8.96 & $2\big[\cos(k_x)+\cos(k_y)+\cos(k_x')+\cos(k_y')\big]$& $4\big[\cos(k_x)+\cos(k_y)\big]$\\
\cline{2-9}
 &  \multirowcell{2}{$\delta t_2 = \delta t_{\mathbf{R},\mathbf{R}^\prime}$}& \multirowcell{2}{(1,0)} & \multirowcell{2}{(1,1)} & \multirowcell{2}{-7.16} & \multirowcell{2}{-6.28} & \multirowcell{2}{-5.47} & \multirowcell{2}{$4\big(\cos(k_x - k_x')\big[\cos(k_y)+\cos(k_y')\big]\Big.$\\+$\;\cos(k_y - k_y')\big[\cos(k_x)+\cos(k_x')\big]\big)$}& \multirowcell{2}{$8\big[\cos(k_x)+\cos(k_y)\big]$}\\
 & & & & & & & &\\
\cline{2-9}
 &  $\delta t_0^\prime = \delta t_{\mathbf{R},\mathbf{R}^\prime}$ & (0,0) & (1,1) & -1.75 & -1.31 & -0.47 & $4\big[\cos(k_x)\cos(k_y) + \cos(k_x')\cos(k_y')\big]$& $8 \cos(k_x)\cos(k_y)$\\

\hline
\end{tabular}
\caption{\label{tab:impuritiesSr} Impurity parameters for the Sr dopant in LSCO. \textit{Ab-initio} potentials from Wannier-projected DFT are tabulated in units of the DFT-derived nearest-neighbor hopping, $|t| = 524.7$~meV, at three hole-doping levels: the undoped La$_{35}$SrCu$_{18}$O$_{72}$ supercell; $p_\mathrm{vH}$, with the van Hove singularity tuned to the Fermi level; and $p_\mathrm{vH} + 7\%$. Representative values of the coordinates $\mathbf{R}$ and $\mathbf{R}^\prime$ are given in a coordinate system where the defect is located a distance $z$ above the origin. For Sr$^\mathrm{near}$ ($z = 1.8$~\AA), the origin is centered on a CuO$_2$ plaquette. For Sr$^\mathrm{far}$ ($z = 4.8$~\AA), the origin is centered on a Cu atom. Additional, symmetry-related copies are implied, and when included produce the momentum-dependent form factors shown, with momenta in units of $1/a$, the inverse lattice parameter.}
\end{table*}

\begin{table*}[t]
\renewcommand{\arraystretch}{1.7}
\begin{tabular}
{|c|c|c|c|c|c|c|}
\hline
\multirowcell{2}{Defect\\\&\\location} & \multirowcell{2}{Impurity\\term} & \multicolumn{2}{c|}{Rep.~coords.} & \multirowcell{2}{Impurity potential \\ $\Delta E/|t| \times 1000\big.$} & \multirowcell{2}{Form factor\\$f(\mathbf{q})$ or $f(\mathbf{k},\mathbf{k}^\prime)\big.$}& \multirowcell{2}{Form factor\\ $f(\mathbf{q} = 0) \equiv f(\mathbf{k} = \mathbf{k}^\prime)\big.$}\\\cline{3-4}
  &  & $\;\;\;\,\mathbf{R}/a\;\;\;\,$ & $\mathbf{R}^\prime/a$ & & & \\ 
  
\hline\hline

 \multirowcell{7}{Cu$^\mathrm{near}$\\$z = 4.7$~\AA\\site-\\centered} 
& $V_0 = V_\mathbf{R}$ & (0,0) & -- & 155.72 & 1 & 1\\ 
\cline{2-7}
 & $V_2 = V_\mathbf{R}$ & (1,0) & -- & 77.50 & $2\big[\cos(q_x) + \cos(q_y)\big]$ & 4\\
\cline{2-7}
 & $V_4 = V_\mathbf{R}$ & (1,1) & -- & 66.97 & $4\;\cos(q_x)\cos(q_y)$ & 4\\
\cline{2-7}
 &  $\delta t_0 = \delta t_{\mathbf{R},\mathbf{R}^\prime}$ & (0,0) & (1,0) & -9.15 & $2\big[\cos(k_x)+\cos(k_y)+\cos(k_x')+\cos(k_y')\big]$ & $4\big[\cos(k_x)+\cos(k_y)\big]$\\
\cline{2-7}
 &  \multirowcell{2}{$\delta t_2 = \delta t_{\mathbf{R},\mathbf{R}^\prime}$}& \multirowcell{2}{(1,0)} & \multirowcell{2}{(1,1)} & \multirowcell{2}{-7.39} & \multirowcell{2}{$4\big(\cos(k_x - k_x')\big[\cos(k_y)+\cos(k_y')\big]\Big.$\\$+\;\cos(k_y - k_y')\big[\cos(k_x)+\cos(k_x')\big]\big)$}& \multirowcell{2}{$8\big[\cos(k_x)+\cos(k_y)\big]$}\\
 & & & & & &\\
\cline{2-7}
 &  $\delta t_0^\prime = \delta t_{\mathbf{R},\mathbf{R}^\prime}$ & (0,0) & (1,1) & -2.88 & $4\big[\cos(k_x)\cos(k_y) + \cos(k_x')\cos(k_y')\big]$& $8 \cos(k_x)\cos(k_y)$\\
 
\hline\hline

\multirowcell{9}{Cu$^\mathrm{far}$\\$z = 6.9$~\AA\\plaquette-\\centered} 
& $V_1 = V_\mathbf{R}$ & $\left(\frac{1}{2},\frac{1}{2}\right)$ & -- & 100.50 & $4\cos\left(\frac{q_x}{2}\right)\cos\left(\frac{q_y}{2}\right)$ & 4\\ 
\cline{2-7}
 & $V_3 = V_\mathbf{R}$ & $\left(\frac{1}{2},\frac{3}{2}\right)$ & -- & 37.29 & $4\Big[\cos\Big(\frac{q_x}{2}\Big)\cos\Big(\frac{3q_y}{2}\Big)+ \cos\Big(\frac{3 q_x}{2}\Big)\cos\Big(\frac{q_y}{2}\Big)\!\Big]$ & 8\\
\cline{2-7}
  &  \multirowcell{2}{$\delta t_1 = \delta t_{\mathbf{R},\mathbf{R}^\prime}$}& \multirowcell{2}{$\left(\frac{1}{2},\frac{1}{2}\right)$} & \multirowcell{2}{$\left(\frac{1}{2},-\frac{1}{2}\right)$} & \multirowcell{2}{-6.45} & \multirowcell{2}{$8\Big[\cos\Big(\frac{k_x + k_y}{2}\Big)\cos\Big(\frac{k_x' - k_y'}{2}\Big)$\\ $+ \cos\Big(\frac{k_x - k_y}{2}\Big)\cos\Big(\frac{k_x' + k_y'}{2}\Big)\Big]$} &  \multirowcell{2}{$8\big[\cos(k_x)+\cos(k_y)\big]$}\\
 & & & & & &\\
\cline{2-7}
  &  \multirowcell{4}{$\delta t_3 = \delta t_{\mathbf{R},\mathbf{R}^\prime}$}& \multirowcell{4}{$\left(\frac{1}{2},\frac{1}{2}\right)$} & \multirowcell{4}{$\left(\frac{1}{2},\frac{3}{2}\right)$} & \multirowcell{4}{-2.83} & \multirowcell{4}{$4\Big[\cos\Big(\frac{2 k_x + 2 k_y - k_x' - k_y'}{2}\Big) 
\cos\Big(\frac{k_x - k_y}{2}\Big)$\\
  $+ \cos\Big(\frac{2 k_x - 2 k_y - k_x' + k_y'}{2}\Big) \cos\Big(\frac{k_x + k_y}{2}\Big)$\\
  $+ \cos\Big(\frac{k_x + k_y - 2 k_x' - 2 k_y' }{2}\Big)  
  \cos\Big(\frac{k_x' - k_y'}{2}\Big)$\\
  $+ \cos\Big(\frac{k_x - k_y -2k_x' + 2 k_y' }{2}\Big) 
  \cos\Big(\frac{k_x' + k_y'}{2}\Big)\Big]$}&  \multirowcell{4}{$8\big[\cos(k_x)+\cos(k_y)\big]$}\\
 & & & & & &\\
 & & & & & &\\
 & & & & & &\\
\cline{2-7}
 &  $\delta t_1^\prime = \delta t_{\mathbf{R},\mathbf{R}^\prime}$ & $\left(\frac{1}{2},\frac{1}{2}\right)$ & $\left(-\frac{1}{2},-\frac{1}{2}\right)$ & -0.89 & $8 \cos\Big(\frac{k_x + k_x' }{2}\Big) \cos\Big(\frac{k_y + k_y'}{2}\Big)$& $8 \cos(k_x)\cos(k_y)$\\
 
\hline
\end{tabular}
\caption{\label{tab:impuritiesTl} Impurity parameters for Cu substituted onto the Tl site in Tl-2201. \textit{Ab-initio} potentials from Wannier-projected DFT are tabulated in units of the DFT-derived nearest-neighbor hopping, $|t| = 537.8$~meV, for a Tl$_{35}$Ba$_{36}$Cu$_{19}$O$_{108}$ supercell. Representative values of the coordinates $\mathbf{R}$ and $\mathbf{R}^\prime$ are given in a coordinate system where the defect is located a distance $z$ above the origin. For Cu$^\mathrm{near}$ ($z = 4.7$~\AA), the origin is centered on a Cu site. For Cu$^\mathrm{far}$ ($z = 6.9$~\AA), the origin is centered on a CuO$_2$ plaquette. Additional, symmetry-related copies are implied, and when included produce the momentum-dependent form factors shown, with momenta in units of $1/a$, the inverse lattice parameter.}
\end{table*}

We note that the impurity potential needs to be partitioned from its super-images. To this end we use the partitioning scheme detailed in the supplement of Ref.~\onlinecite{Berlijn2011}. The Wannier-function-based Hamiltonian has been derived using the \mbox{Wannier90} software. \citep{Mostofi_2008} Specifically, we projected \mbox{Cu-$d_{x^2-y^2}$} orbitals on the bands within low energy windows. These low energy windows were taken to be $[-4.5,2.5]$~eV for the La$_4$Cu$_2$O$_8$ normal cell and La$_{35}$SrCu$_{18}$O$_{72}$ supercell; $[-4.6,2.3]$~eV for the La$_{36}$Cu$_{18}$O$_{71}$ supercell;  and $[-3,2.8]$~eV for the Tl$_4$Ba$_4$Cu$_2$O$_{12}$ normal cell and the Tl$_{35}$Ba$_{36}$Cu$_{19}$O$_{108}$ supercell. To obtain projected Wannier functions we set $\mathtt{num\_iter}=0$. Due to the large number of orbitals in the supercells, the disentanglement procedure became unstable. Therefore, we set $\mathtt{dis\_num\_iter}=0$. As a result, the Wannier bands in some parts of $k$-space were shifted downwards in energy relative to the DFT bands (see Fig.~\ref{fig:band_comparison}). The most important consideration for the purposes of calculating impurity potentials is that the underlying Wannier functions have well-localized \mbox{Cu $d_{x^2 - y^2}/$O $p_{x,y}$} anti-bonding character, as can be seen in Fig.~\ref{fig:band_comparison}(c). In any case, some of the mismatch is cancelled because the impurity potential is the difference between the single impurity supercell Hamiltonian and the reference Hamiltonian, as shown in Eq.~\ref{eqn:Himp}.

Results for the \textit{ab-initio} impurity-potential calculations are presented in \mbox{Tables~\ref{tab:impuritiesAOV}--\ref{tab:impuritiesTl}}, with numerical values of the real-space matrix elements \mbox{$\delta H_{\mathbf{R}\mathbf{R}^\prime}^i = V_\mathbf{R}^i \delta_{\mathbf{R},\mathbf{R}^\prime}~\mathrm{and }~\delta t_{\mathbf{R}\mathbf{R}^\prime}^i$} given for representative choices of $\mathbf{R}$ and $\mathbf{R}^\prime$, in units of the DFT-derived nearest-neighbor hopping $|t|$.  As shown in Fig.~\ref{fig:potential_schematic}, it is useful to group together  terms that are equivalent by symmetry.  Transforming to momentum space, as shown in Eq.~(\ref{eqn:matrixelements}), the grouping of like terms leads to a set of form factors that enter the matrix elements $V_{\veck,\veckP}^i$.  For the site energies these are
\begin{equation}
    f(\veck,\veckP) = f(\mathbf{q}) = \!\!\sum_{\{\mathrm{equiv.}\,\mathbf{R}\}}e^{i(\veck - \veckP)\cdot \mathbf{R}} = \!\!\sum_{\{\mathrm{equiv.}\,\mathbf{R}\}}e^{i\mathbf{q}\cdot \mathbf{R}}\;,
\end{equation}
and for the hopping modifications
\begin{equation}
    f(\veck,\veckP) =  \!\!\!\!\sum_{\{\mathrm{equiv.}\,(\mathbf{R},\mathbf{R}^\prime )\}}e^{i \veck \cdot \mathbf{R}} e^{-i \veckP \cdot \mathbf{R}^\prime}\;,
\end{equation}
where in each case the phase factors are summed over the set of equivalent lattice vectors $\mathbf{R}$ or pairs of lattice vector $(\mathbf{R},\mathbf{R}^\prime)$.  By using a 2D coordinate system in which the impurity sits directly above (or below) the origin, we obtain form factors that are purely real.  These are listed beside their corresponding impurity terms in \mbox{Tables~\ref{tab:impuritiesAOV}--\ref{tab:impuritiesTl}}, along with the form factor evaluated in the limit $\veck = \veckP$ (i.e., $\mathbf{q} = 0$), as this is the form in which it enters the first-order self energy. In this limit, the site-energy form factors are constants that can be absorbed into the chemical potential.  Interestingly, in the $\mathbf{q} = 0$ limit, the form factors derived from the hopping modifications  take on precisely the form of the band dispersions they modify, meaning that they can be be directly absorbed into a renormalization of the hopping integrals.  In this sense, we have taken the first-order self energy into account from the outset, since  we base our energy dispersions on ARPES-derived tight-binding parameterizations.

To account for the doping dependence of the \textit{ab-initio} impurity potentials in LSCO, linear interpolation is used between $p_\mathrm{vH}$ and $p_\mathrm{vH} + 7\%$. The overdoped regime is sufficiently far removed from the undoped compounds that interpolation does not make sense below $p_\mathrm{vH}$; constant extrapolation of the $p_\mathrm{vH}$ values is used instead.

\section{Momentum sums and the van Hove singularity}
\label{appendix_van_Hove}

In calculations such as those carried out here, Brillouin-zone momentum sums are typically converted to Fermi-surface integrals, by linearizing the dispersion $\xi_\veck$ in the vicinity of the Fermi surface: i.e., \mbox{$\xi_\veck \to \hbar |\mathbf{v}_{F,k_\parallel}|\delta k_\perp$}. In a 2D system of area $L^2$,
\begin{align}
    \sum_\veck^\mathrm{BZ} A_\veck & \to \frac{L^2}{(2 \pi)^2} \int_\mathrm{FS}\!\!\!\mathrm{d} k_\parallel \!\!\int\!\mathrm{d} k_\perp A(k_\parallel,k_\perp)\\
    &\approx \frac{L^2}{\hbar (2 \pi)^2} \int_\mathrm{FS}\!\!\!\mathrm{d} k_\parallel \!\! \int_{-\infty}^\infty\!\!\!\!\!\mathrm{d} \xi \frac{1}{| \mathbf{v}_{F,k_\parallel} |} A(k_\parallel,\xi)\;,
\end{align}
with the $\xi$ integral carried out in closed form, and the remaining $k_\parallel$ integral expressed as a Fermi surface average.  Implicit in this conversion is that the Fermi velocity $\mathbf{v}_{F,k_\parallel}$ not go to zero anywhere on the Fermi surface, and that the kernel of the integral, $A(k_\parallel,\xi)$, die off fast enough that the $\xi$ integral  not extend beyond the boundary of the irreducible Brillouin zone.  

\begin{figure*}[t]
    \centering
        \includegraphics[width=\linewidth,scale=1.0]{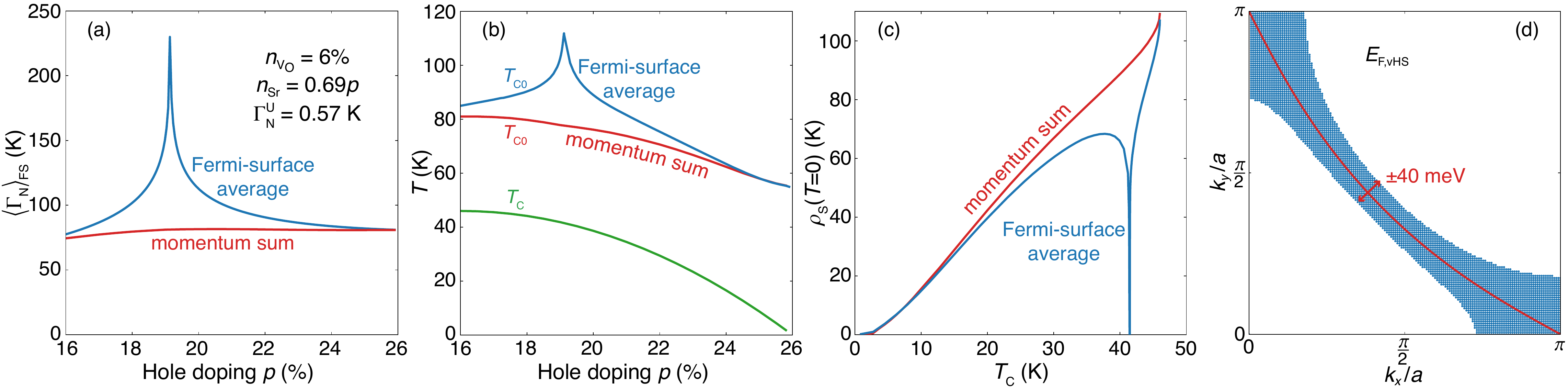}
    \caption{Comparison of Fermi-surface-average and momentum-sum methods in LSCO, in the vicinity of the van Hove singularity (vHS) at hole doping $p = 19\%$. (a) Average normal-state scattering rate, $\langle \Gamma_N \rangle_\mathrm{FS}$. (b) Clean-limit transition temperatures $T_{c0}(p)$ required to produce the parabolic dome of transition temperature, $T_c(p)$. (c) Resulting zero-temperature superfluid density, $\rho_s(T = 0)$. (d) One quadrant of the 2D Brillouin zone, at the Lifshitz transition ($p = 19\%$), schematically showing how the momentum sum (discrete points) can be implemented} using an energy cutoff of $\pm 40$~meV about the Fermi surface (solid line).  The plots in (a)--(c) use the disorder parameters shown in panel (a).
    \label{fig:vHS}
\end{figure*}

These requirements fail spectacularly in LSCO, which undergoes a Lifshitz transition at $p \approx 19\%$ when the van Hove singularity (vHS) passes through the Fermi energy, as shown in Fig.~\ref{fig:vHS}(d). For our realistic disorder model, in which the calculated scattering rates and self-energies (e.g., in Eqs.~(\ref{eqn:tau0_self_energy}) and (\ref{eqn:tau1_self_energy})) are highly sensitive to the density of final states, the consequences of persisting with Fermi surface averages can be dire.  This is seen in Figs.~\ref{fig:vHS}(a)--(c) for LSCO.  When the calculation is restricted to the Fermi surface, the normal-state scattering rate $\langle \Gamma_N \rangle_\mathrm{FS}$ diverges at the vHS; the required pairing strength, parameterized by the clean-limit transition temperature $T_{c0}(p)$, has a sharp peak; and the zero-temperature superfluid density, $\rho_s(T=0)$, is driven to zero. None of these effects is physical or in accord with experiment. By contrast, when the momentum-sum method is used, $\langle \Gamma_N \rangle_\mathrm{FS}$, $T_{c0}$ and $\rho_s(T=0)$ pass smoothly and monotonically through the vHS. Well away from the vHS, the Fermi-surface calculations agree with the momentum-sum method but, as Figs.~\ref{fig:vHS}(a)--(c) show, the anomalies associated with the vHS persist over a surprisingly wide range, so caution is warranted.

Momentum sums are computationally more expensive than Fermi-surface averaging.  To mitigate this, after taking advantage of symmetry to down-fold all sums into the irreducible octant, an energy cutoff can be implemented to eliminate unneeded $k$-points, on the basis that the kernels of the sums fall off sufficiently rapidly with $\xi_\veck$. We have verified this by testing for convergence, with the result that an energy cutoff of 40~meV safely eliminates any cutoff dependence. In calculations of the sort presented here, the coarse-grained and energy-cut-off momentum sum typically consists of 4700~\mbox{$k$-points} within the irreducible octant. Figure~\ref{fig:vHS}(d) illustrates the situation in LSCO, showing how the $\veck$-sum method nicely regularizes the approach to the vHS at the $(\pi,0)$ point.  However, for completeness, the energy cutoff was removed for the final calculations presented in the paper, which were carried out using full-Brillouin-zone sums consisting of approximately 12000 points.

\section{Effect of V$_\mathbf{O}$ and Sr concentration in LSCO}

\label{appendix_strontium}

\begin{figure}[t]
    \centering
        \includegraphics[width=0.7\columnwidth,scale=1.0]{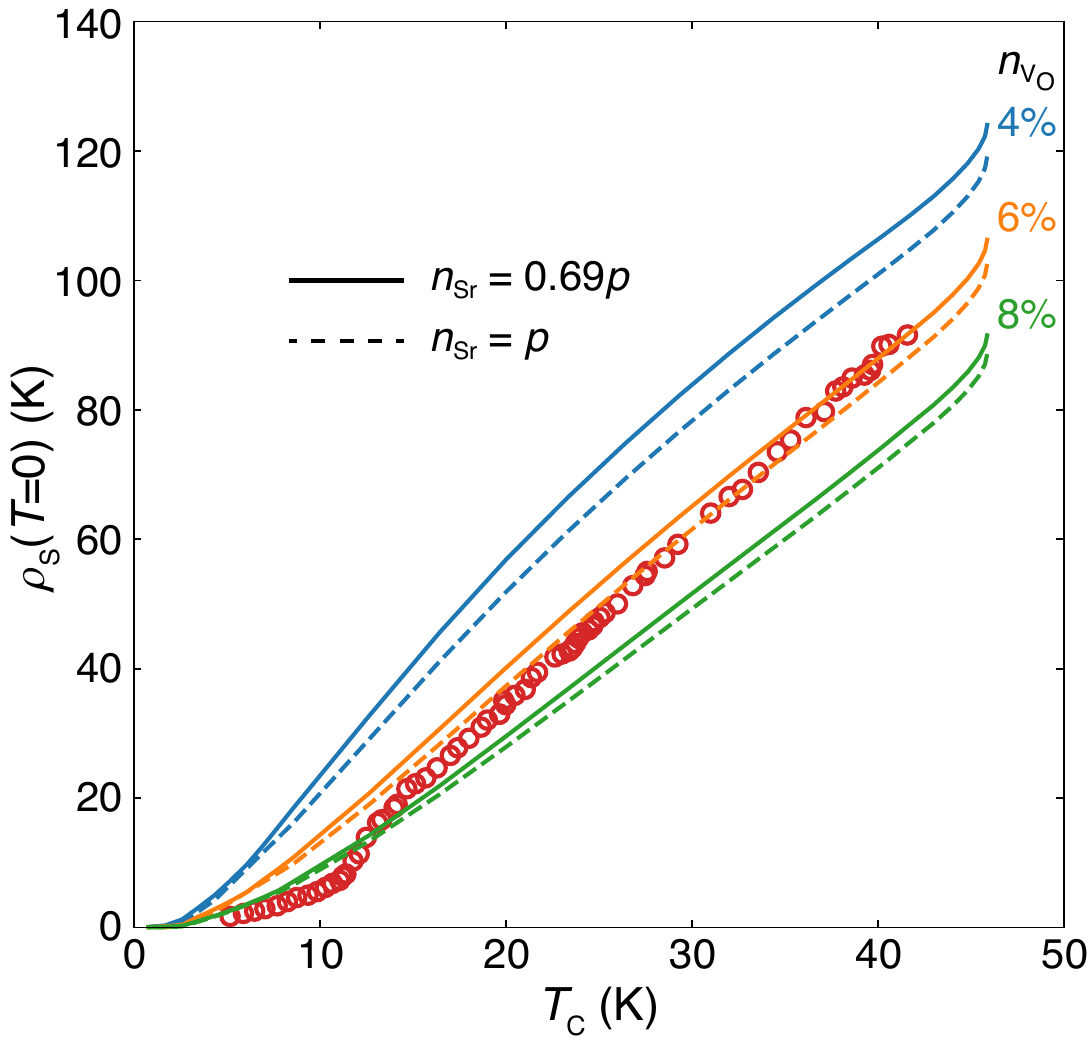}
    \caption{The effect of Sr and apical-oxygen-vacancy concentration on the superfluid density of LSCO. Solid lines are calculated using the doping relation from ARPES Fermi volume, \mbox{$n_\mathrm{Sr} = 0.69 p$}; dashed lines are based on the conventional doping dependence, \mbox{$n_\mathrm{Sr}= p$}. \mbox{$n_{\mathrm{V}_\mathrm{O}} \!= 4, 6$~and 8\%} and $\Gamma_N^u = 0.57$~K. Experimental data from Ref.~\onlinecite{Bozovic:2016ei}.}
    \label{fig:strontiumLSCO}
\end{figure}

The effect of Sr dopant and apical-oxygen-vacancy concentration on the zero-temperature superfluid density of LSCO is shown in Fig.~\ref{fig:strontiumLSCO}.  Two doping dependences of the Sr concentration are compared: one derived from ARPES Fermi volume, \mbox{$n_\mathrm{Sr} = 0.69 p$}; and the conventional relation, \mbox{$n_\mathrm{Sr}= p$}. The impurity potential of the Sr dopants, which is extended in real space and therefore inherently forward scattering, does not have a strong pair-breaking effect, with the superfluid density relatively insensitive to changes in Sr concentration.  Results are also shown for three different concentrations of apical oxygen vacancy.  This defect, which has a nearly pointlike impurity potential, causes significant pair breaking.

\section{Effect of intermediate-strength scatterers}

\label{appendix_intermediate_scatterers}

As mentioned in Sec.~\ref{sec:discussion}, a possible shortcoming of the approach taken in the current paper is that we have treated the out-of-plane dopant impurities in the Born approximation, where the perturbation series for their self energies, Eqs.~(\ref{eqn:tau0_self_energy}) and (\ref{eqn:tau1_self_energy}), are truncated at second order.  This makes the calculation of vertex corrections tractable, and becomes exact in the weak-scattering limit, $V_\mathrm{imp} N_0 \ll 1$. A better approach would be to study these impurities in the $t$-matrix approximation where, by summing the perturbation series to all orders, arbitrary scattering strengths could be accurately accounted for.  This is unfortunately beyond the scope of the current paper, due to the computational cost of implementing  vertex corrections in the full $t$-matrix approximation.

\begin{figure}[t]
    \centering
        \includegraphics[width=0.7\columnwidth,scale=1.0]{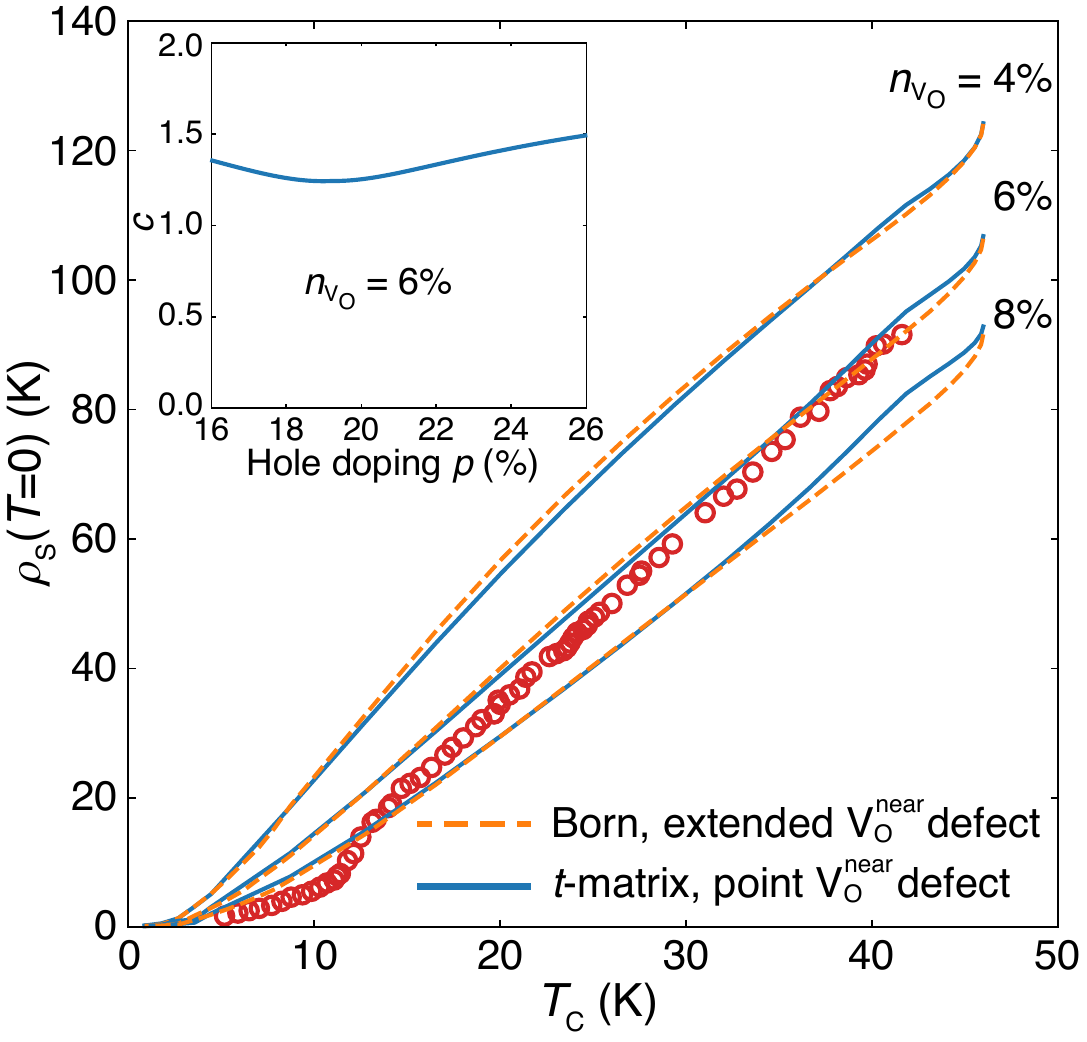}
    \caption{Zero-temperature superfluid density for LSCO, calculated by treating the near apical oxygen vacancy, $V_\mathrm{O}^\mathrm{near}$, alternately in the $t$-matrix and Born approximations.  In the $t$-matrix approximation (solid lines), $V_\mathrm{O}^\mathrm{near}$ is modelled by an intermediate-strength point scatterer of scattering parameter $c$.  In the Born approximation (dashed lines), the full extended impurity potential \mbox{$n_{\mathrm{V}_\mathrm{O}} \!= 4, 6$~and 8\%}, \mbox{$n_\mathrm{Sr} = 0.69 p$} and $\Gamma_N^u = 0.57$~K.     Experimental data from Ref.~\onlinecite{Bozovic:2016ei}. Inset: $c = \cot \delta$ for the $V_\mathrm{O}^\mathrm{near}$ impurity, for $n_{\mathrm{V}_\mathrm{O}} \!= 6\%$.
    } 
    \label{fig:intermediateLSCO}
\end{figure}

To test the adequacy of the Born approximation, we focus on the strongest out-of-plane scatterer in LSCO, the near apical oxygen vacancy, $V_\mathrm{O}^\mathrm{near}$.  As can be seen in Table~\ref{tab:impuritiesAOV}, the impurity potential for $V_\mathrm{O}^\mathrm{near}$ is nearly pointlike, with \mbox{$V_\mathrm{imp} = V_0 \gg V_2, V_4, ...$}. To good approximation this enables us to treat the $V_\mathrm{O}^\mathrm{near}$ defects as point scatterers, and therefore to use the $t$-matrix approximation. The contribution of $V_\mathrm{O}^\mathrm{near}$ to the $\tau_0$ self energy in the point-scattering limit then takes the form
\begin{equation}
    \Sigma_0^{t-\mathrm{matrix}} = \Gamma \frac{G_0}{c^2 + G_0^2}\;,
    \label{eqn:t-matrix_self_energy}
\end{equation}

\noindent where $\Gamma =  n_{\mathrm{V}_\mathrm{O}}/(\pi N_0)$ parameterizes the concentration of the apical oxygen vacancies, and $c = 1/(\pi V_\mathrm{imp}N_0)$ is the cotangent of the scattering phase shift. 

To implement the $t$-matrix approximation for the near apical oxygen vacancies, the Born term for $V_\mathrm{O}^\mathrm{near}$ in Eq.~(\ref{eqn:tau0_self_energy}) is replaced by Eq.~(\ref{eqn:t-matrix_self_energy}). Note that for point scatterers there is no contribution to explicit gap renormalization in Eq.~(\ref{eqn:tau1_self_energy}), and no contribution to vertex corrections.  Figure~\ref{fig:intermediateLSCO} shows the results for zero-temperature superfluid density in LSCO. Born and $t$-matrix approximations are in very close agreement over the whole doping range, with a small deviation in the vicinity of the van Hove doping.  The doping dependence of the $V_\mathrm{O}^\mathrm{near}$ scattering parameter, plotted in the inset of Fig.~\ref{fig:intermediateLSCO}, shows the reason for the deviation: $c(p)$ has a small dip near $p = 19\%$ as a result of the van-Hove enhancement of $N_0$ at this doping.

Referring to Table~\ref{tab:impuritiesSr}, we see that the impurity potential for Sr$^\mathrm{near}$ is significantly weaker than for $V_\mathrm{O}^\mathrm{near}$, even taking into account that the plaquette-centered Sr$^\mathrm{near}$ affects its nearest four Cu neighbors equally.  We are therefore confident that although the out-of-plane defects approach intermediate scattering strength ($c \sim 1$), the Born approximation remains valid.

\begin{widetext}
\section{Disorder renormalization of 2-particle properties}
\label{appendix_vertex_corrections}
Within the ladder approximation, the current vertex $\mathbf{\Lambda}(\veck, \omega_n)$ can be approximated by a sum of ladder diagrams and then resolved into components $\gamma_0$, $\gamma_1$ and $\gamma_3$ in \mbox{particle--hole} space.\cite{Skalski:1964,Durst:2000}

\begin{align} \label{eqn:ladderApprox}
    \mathbf{\Lambda}(\veck, \omega_n) &= \mathbf{v}_{\veck} \tau_{0}
+ \frac{1}{N} \sum_{\veckP} \sum_i n_i\left|V_{\veck,\veckP}^i\right|^2 \textbf{v}^j_{\veckP}\tau_{3} \uline{G}(\veckP\!,\omega_n) \mathbf{\Lambda}(\veckP\!, \omega_n) \uline{G}(\veckP\!,\omega_n) \tau_{3}\\
\gamma_0(\veck,\omega_n) &= 1 + \frac{1}{N}\sum_{\veckP} \sum_{i} s_{\veck,\veckP} n_i \left|V_{\veck,\veckP}^i\right|^2 \frac{\gamma_0(\veckP\!, \omega_n) \left( \tilde\Delta_{\veckP\!\!,n}^2 + \xi_{\veckP}^2 -\tilde\omega_{\veckP\!\!,n}^2\right) - 2\gamma_1(\veckP\!, \omega_n) \tilde\Delta_{\veckP\!\!,n} \tilde\omega_{\veckP\!\!,n} +  2i\gamma_3(\veckP\!, \omega_n) \xi_{\veckP}\tilde\omega_{\veckP\!\!,n}}{\left(\tilde\Delta_{\veckP\!\!,n}^2 + \xi_{\veckP}^2 + \tilde\omega_{\veckP\!\!,n}^2\right)^2}\\
\gamma_1(\veck,\omega_n) &= \frac{1}{N}\sum_{\veckP} \sum_{i} s_{\veck,\veckP} n_i \left|V_{\veck,\veckP}^i\right|^2
\frac{-2\gamma_0(\veckP\!, \omega_n)\tilde\Delta_{\veckP\!\!,n}\tilde\omega_{\veckP\!\!,n} + \gamma_1(\veckP\!, \omega_n) \left(-\tilde\Delta_{\veckP\!\!,n}^2 + \xi_{\veckP}^2 + \tilde\omega_{\veckP\!\!,n}^2\right) + 2i\gamma_3(\veckP\!, \omega_n) \xi_{\veckP}\tilde\Delta_{\veckP\!\!,n}}{\left(\tilde\Delta_{k'}^2 + \xi_{\veckP}^2 + \tilde\omega_{\veckP\!\!,n}^2\right)^2}\\
\gamma_3(\veck,\omega_n) &= \frac{1}{N}\sum_{\veckP} \sum_{i} s_{\veck,\veckP} n_i \left|V_{\veck,\veckP}^i\right|^2
\frac{2i \gamma_0(\veckP\!, \omega_n)\xi_{\veckP} \tilde\omega_{\veckP\!\!,n}  +  2i \gamma_1(\veckP\!, \omega_n) \tilde\Delta_{\veckP\!\!,n}\xi_{\veckP}  - \gamma_3(\veckP\!, \omega_n) \left(\tilde\Delta_{\veckP\!\!,n}^2 - \xi_{\veckP}^2 + \tilde\omega_{\veckP\!\!,n}^2\right)}{\left(\tilde\Delta_{\veckP\!\!,n}^2 + \xi_{\veckP}^2 + \tilde\omega_{\veckP\!\!,n}^2\right)^2}
\label{eq:vertexFunctions}
\end{align}
where $s_{\veck, \veckP} = \mathbf{v}_{\veckP}^j \!\cdot\mathbf{v}_{\veck}^j/|\mathbf{v}_{\veck}^j|^2$. When it is safe to assume a linearized spectrum $\xi_\veck$, the $\xi$ integrations in Eqs.~\ref{eqn:ladderApprox} to \ref{eq:vertexFunctions} can be carried out, allowing the vertex functions to be recast as Fermi surface averages.
    \begin{align}\label{eqn:ladderApproxFS}
    \mathbf{\Lambda}(\veck, \omega_n) & = \mathbf{v}_{\veck} \tau_{0} 
    + N_0 \left\langle \sum_i n_i \left|V^i_{\veck,\veckP}\right|^2\textbf{v}_{\veckP}^j \tau_{3} \uline{G}(\veckP\!,\omega_n) \mathbf\Lambda (\veckP\!, \omega_n) \uline{G}(\veckP\!, \omega_n) \tau_{3} \right\rangle\FSavgP\\
    \gamma_{0}(\veck,\omega_n) & = 1
    + N_0\;\Bigg\langle \sum_i n_i \left|V_{\veck,\veckP}^i\right|^2 s_{\veck,\veckP} \frac{\tilde{\Delta}_{\veckP\!\!,n}^{2}\gamma_{0}(\veckP\!, \omega_n) - \tilde{\Delta}_{\veckP\!\!,n} \tilde{\omega}_{\veckP\!\!,n}\gamma_{1}(\veckP\!, \omega_n)}{\left(\tilde{\omega}^2_{\veckP\!\!,n} +\tilde{\Delta}_{\veckP\!\!,n}^{2}\right)^{3/2}} \Bigg\rangle\FSavgP \\
    \gamma_{1}(\veck, \omega_n) & = - N_0\;\Bigg\langle \sum_i n_{i}  \left|V_{\veck,\veckP}^i\right|^2 s_{\veck,\veckP} \frac{\tilde{\Delta}_{\veckP\!\!,n} \tilde{\omega}_{\veckP\!\!,n}\gamma_{0}(\veckP\!,\omega_n) - \tilde{\omega}_{\veckP\!\!,n}^{2}\gamma_{1}(\veckP\!,\omega_n)}{\left(\tilde{\omega}^2_{\veckP\!\!,n} +\tilde{\Delta}_{\veckP\!\!,n}^{2}\right)^{3/2}} \Bigg\rangle\FSavgP \\
    \gamma_{3}(\veck, \omega_n) & = 0\;.
    \end{align}
\end{widetext}


%

\end{document}